\documentclass[%
reprint,
superscriptaddress,
frontmatterverbose,
showpacs,preprintnumbers,
 amsmath,amssymb,
 aps,pra,Ifloatfix,
citeautoscript,
]{revtex4-2}

\usepackage{mdframed}

\usepackage[pdftex]{graphicx}

\usepackage{mathrsfs}
\usepackage[linkcolor=blue,colorlinks=true,citecolor=blue,filecolor=blue]{hyperref}
\usepackage{amsmath}
\usepackage[english]{babel}
\usepackage{booktabs}
\usepackage{amssymb}
\usepackage{type1cm}
\usepackage{braket}
\usepackage{tikz}
\usepackage{pgfplots}
\usepackage{lipsum}
\usepackage{xcolor}
\usepackage{graphicx}
\usepackage{subcaption}
\usepackage{dcolumn}
\usepackage{bm}
\usepackage{physics}
\usepackage[utf8]{inputenc}
\usepackage{multirow}
\usepackage{tikz}

\definecolor{Midnight_Blue}{rgb}{0.1, 0.1, 0.6}

\usepackage{cleveref}
\crefname{equation}{Eq.}{Eqs.}
\Crefname{equation}{Equation}{Equations}
\crefname{figure}{Fig.}{Figs.}
\Crefname{figure}{Figure}{Figures}
\crefname{figure}{Fig.}{Figs.}
\Crefname{figure}{Figure}{Figures}
\crefname{section}{Supplemental Material Section}{Supplemental Material Sections}
\Crefname{section}{Supplemental Material Section}{Supplemental Material Sections}
\crefname{appendix}{Appendix}{Appendices}
\Crefname{appendix}{Appendix}{Appendices}
\crefname{table}{Table}{Tables}
\Crefname{table}{Table}{Tables}

\usepackage{enumitem,amssymb}
\newlist{todolist}{itemize}{2}
\setlist[todolist]{label=$\square$}
\usepackage{pifont}

\usepackage{etoolbox}
\pgfplotsset{compat=1.18} 
\begin{document}

\title{Noise-reduction of Multimode Gaussian Boson Sampling Circuits via Unitary Averaging}

\author{S. Nibedita Swain}
\email{swain.snibedita@gmail.com}
\affiliation{School of Mathematical and Physical Sciences,
University of Technology Sydney, Ultimo, NSW 2007, Australia}

\affiliation{Sydney Quantum Academy, Sydney, NSW 2000, Australia}
\affiliation{Centre for Quantum Computation and Communication Technology, School of Mathematics
and Physics, University of Queensland, Brisbane, Queensland 4072, Australia}
\author{Ryan J. Marshman}
\affiliation{Centre for Quantum Computation and Communication Technology, School of Mathematics
and Physics, University of Queensland, Brisbane, Queensland 4072, Australia}

\author{Alexander S. Solntsev}
\affiliation{School of Mathematical and Physical Sciences,
University of Technology Sydney, Ultimo, NSW 2007, Australia}

\author{Timothy C. Ralph}
\affiliation{Centre for Quantum Computation and Communication Technology, School of Mathematics
and Physics, University of Queensland, Brisbane, Queensland 4072, Australia}

\begin{abstract}
   We improve Gaussian Boson Sampling (GBS) circuits by integrating the unitary averaging (UA) protocol, previously demonstrated to protect unknown Gaussian states from phase errors [Phys. Rev. A 110, 032622]. Our work extends the applicability of UA to mitigate arbitrary interferometric noise, including beam-splitter and phase-shifter imperfections. Through comprehensive numerical analysis, we demonstrate that UA consistently achieves higher fidelity and success probability compared to unprotected circuits, establishing its robustness in noisy conditions. Remarkably, enhancement is maintained across varying numbers of modes with respect to the noise. We further derive a power-law formula predicting performance gains in large-scale systems, including 100-mode and 216-mode configurations. A detailed step-by-step algorithm for implementing the UA protocol is also provided, offering a practical roadmap for advancing near-term quantum technologies.
\end{abstract}

\date{\today}

\frenchspacing

\maketitle

\section{Introduction}
Optical quantum systems can be employed in various quantum technologies, including computing and communication, offering distinct advantages \cite{BAC19}. Current practical applications often rely on Gaussian source states, such as squeezed states \cite{weedbrook2012gaussian, lvovsky2015squeezed}, and linear networks, enabling tasks like continuous variable teleportation \cite{pirandola2015advances}, quantum key distribution \cite{xu2020secure}, and Gaussian Boson sampling (GBS) \cite{lund2014boson,Hamilton2017_GBS,ZHO20,rahimi2015can,quesada2022quadratic}. However, achieving scalable quantum applications requires effective noise control strategies. Since deterministic linear optical processing is non-universal, traditional error correction methods are not directly applicable, necessitating alternative approaches.

Unitary averaging (UA) \cite{marshman2018passive} is an alternative framework that has proven effective in mitigating phase errors within continuous-variable (CV) linear optical systems \cite{swain2024improving}. 
These setups involve Gaussian source states being injected into single-mode linear networks, where a single copy of an unknown state is transmitted through multiple channels before being non-deterministically recombined. These findings demonstrated that Gaussian encoding combined with vacuum projection—used during the recombination process—offers significant advantages in guarding Gaussian states from phase errors. Notably, this protocol maintains its effectiveness when optical losses are present.

In this paper, we extend the UA protocol to multimode Gaussian systems, incorporating linear optical interferometer noise, such as fluctuations of beamsplitter (BS) and phaseshifter (PS) parameters. We consider set-ups similar to those found in Gaussian boson sampling (GBS) \cite{Hamilton2017_GBS}. BS and PS noise typically arises due to small fluctuations in path length
and timing between the optical modes. This may be between the quantum modes themselves or the quantum modes and a reference mode (the local oscillator). Through out, we refer to linear optical interferometer noise simply as `noise'.
In GBS circuits, a multimode Gaussian state is prepared and subsequently measured using photon detectors. This Gaussian state is typically generated by passing squeezed light through a linear-optical interferometer, although more advanced versions may include displacements alongside squeezing operations. GBS circuits can be used for state production as well as non-universal quantum computation.
As a model of quantum computation, GBS is one of only a few architectures to have demonstrated a quantum computational advantage — the ability of quantum devices to surpass classical computers in solving specific computational tasks \cite{aghaee2025scaling}. State-of-the-art experiments now feature systems with up to 216 modes \cite{zhong2021phase,madsen2022quantum}.

Here, after a brief review, we begin by applying UA to small scale GBS circuits and demonstrate significant improvements, achieved with high probability of success. We then identify various scaling rules for fidelity and success probability which are supported by numerical tests. We then apply these results to large-scale circuits and present evidence that BS and PS noises in large-scale GBS computation can be reduced. Finally we introduce a figure of merit for the enhancement seen with UA in GBS type circuits and interestingly find that it is constant with the number of modes with respect to the noise across a large range of unitaries. 

Our results indicate that UA is an effective and practical technique for controlling circuit noise in large scale GBS systems. Although we don't explicitly consider it here, previous work has demonstrated that UA remains effective in the presence of loss \cite{swain2024improving} and is compatible with overarching fault tolerant constructions \cite{PhysRevA.109.062436}.

\section{Unitary Averaging in a nutshell}
\begin{figure}[!htb]
	\centering
		 \includegraphics[width=0.6\linewidth]{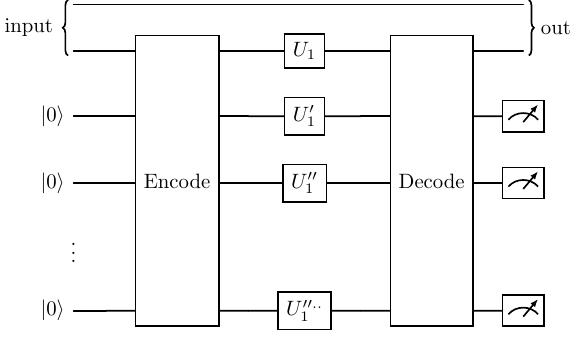}
		\caption{  A passive unitary averaging scheme utilises a beam splitter network for redundant encoding over $n$ modes. One mode of a two-mode state is evenly distributed across $n$ transmission modes. Each mode undergoes independent single-mode unitary noise. The decoding network reverses the encoding process, and upon heralding $n-1$ error modes in the vacuum state, the output state exhibits reduced noise.
  \label{CVUAcir}}
\end{figure}


We begin by reviewing previous work on Gaussian systems, in which the impact of UA on a continuous-variable (CV) system is modelled using a two-mode squeezed vacuum state, as described in \cite{swain2024improving}.
In this model, a passive beam splitter encoding network is employed to place the input state into an equal superposition of $n$ spatial modes. Each of these modes undergo independent transformations, previously simply transmission though phase noise, we are now considering arbitrary linear optical unitary transformations. The decoding network then inverts the encoding such that each superposition should constructively interfier in the top most modes. The remaining `error modes' are then heralded in the vacuum state, such that the heralded output state exhibits suppressed noise. The behaviour is characterised by the resulting output squeezing, purity, and entanglement. 
The two-mode squeezed vacuum state was chosen for characterising the system because it can effectively represent various input states through projective measurement. 
Fig.\ref{CVUAcir} illustrates the circuit discussed in Ref. \cite{swain2024improving}. Below, we provide a brief summary of the corresponding results.

The output state from the the protocol is,

    \begin{align}
    \ket{\psi_{\text{out}}} &= \frac{1}{\mathbf{N}} \left(\cosh{r}\right)^{-1}\sum_{N} (-1)^{N} [e^{i \phi_{\beta}} \tanh{r'} ]^{N} \ket{N, N}\nonumber\\
    &=\hat{S}\left(\chi'\right)\ket{0}\ket{0}
\end{align}
where $\chi'=r'e^{i\phi_{\beta}}$ and the phase terms $\phi_j$ vary randomly and independently around some mean value. We define $\alpha e^{i\phi_{\beta}}= \frac{e^{ i \phi_{1}} + e^{ i \phi_{2}} +...+e^{ i \phi_{n}}}{n}$ and $\tanh{r'} = \alpha\tanh{r}$ with normalisation constant $\mathbf{N}  = \frac{\cosh{r'}}{\cosh{r}}$ and 
probability of success  $P = |\mathbf{N}|^{2}$. 

In the covariance matrix form we have,
\begin{equation}
     \Sigma_{\text{out}(4 \cross 4)} = \left\langle\begin{pmatrix}
          A  &  C  \\
          C^{T} &  B
     \end{pmatrix}\right\rangle \label{eq:covariance matrix}
 \end{equation}
where,
\begin{align}
    A = B 
    & = \begin{pmatrix}
        \frac{1+ \tanh{r^{'}}^{2}}{1 - \tanh{r^{'}}^{2}} & 0\\
        0 &  \frac{1+ \tanh{r^{'}}^{2}}{1 - \tanh{r^{'}}^{2}}
    \end{pmatrix} \label{aeq}
\end{align}

\begin{align}
    C   
    & = \begin{pmatrix}
       - \frac{2\tanh{r^{'}}}{1 - \tanh{r^{'}}^{2}}\cos\left(\phi_{\beta}\right) & \mathscr{C}\\
 \mathscr{C} & \frac{2\tanh{r^{'}}}{1 - \tanh{r^{'}}^{2}}\cos\left(\phi_{\beta}\right) \label{ceq}
    \end{pmatrix}
\end{align}
%
\begin{equation}
\mathscr{C} =   2(\cosh{r'})^{-2} \sum_{N} (N+1) (\tanh{r'} )^{2N+1}\sin\left(\phi_\beta\right) \label{curly c}
\end{equation}
and $C = C^{T}$. 
The analytical expression for the unitary averaging model in the low noise limit assuming the individual random phases $\phi_{j}$ are independent Gaussian parameters with mean zero and variance $v$ can be approximated as
\begin{align}
    \left\langle\tanh{r'}\right\rangle \approx& \Big(1 - \big(\frac{v}{2} - \frac{v}{2n}\big)\Big)\tanh{r} \\
    \left\langle\cos\left(\phi_{\beta}\right)\right\rangle \approx& \cos{\left(\sqrt{\frac{v}{n}}\right)}
\end{align}
where we have assumed $v\ll1$ and
\begin{align}
\left\langle\frac{2\tanh{r^{'}}}{1 - \tanh{r^{'}}^{2}}\cos\left(\phi_{\beta}\right)\right\rangle & \approx \frac{2\left\langle\tanh{r^{'}}\right\rangle}{1 - \left\langle\tanh{r^{'}}\right\rangle^{2}}\left\langle\cos\left(\phi_{\beta}\right)\right\rangle\\
\langle \mathscr{C} \rangle & \approx 0
\end{align}


In summary, we introduced a unitary averaging (UA) scheme with continuous variables that incorporates vacuum detection \cite{swain2024improving}. This approach effectively mitigates the impact of phase noise within the channel. Interestingly, our results demonstrated more than an order of magnitude reduction in noise, along with improvements in squeezing, purity, and entanglement levels in practical systems. Moving forward, we will extend the UA scheme to a multi-mode channel with arbitrary interferometer noise. In the following section, we begin by examining the two-mode channel.

\section{ Multi-mode System}
\subsection{Two-mode System}

\begin{figure*}[!htb]
	\centering
		\begin{subfigure}{\columnwidth}
             \caption{ \label{fig:2MODE1}}
			\includegraphics[width=1.0\columnwidth]{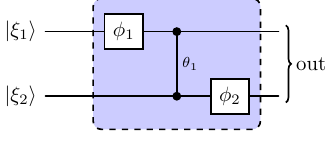}
			
		\end{subfigure}
  		\begin{subfigure}{\columnwidth}
              \caption{ \label{fig:2MODE2}}
			\includegraphics[width=1.0\columnwidth]{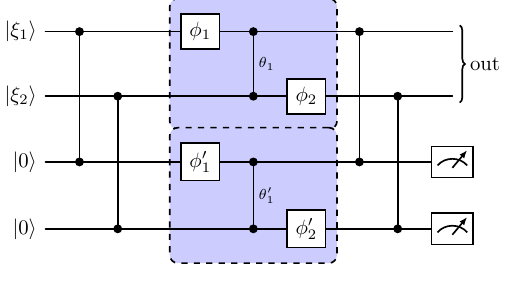}
		\end{subfigure}
        \begin{subfigure}{\columnwidth}
  \caption{\label{fig:2MODE4}}
			\includegraphics[width=1.2\columnwidth]{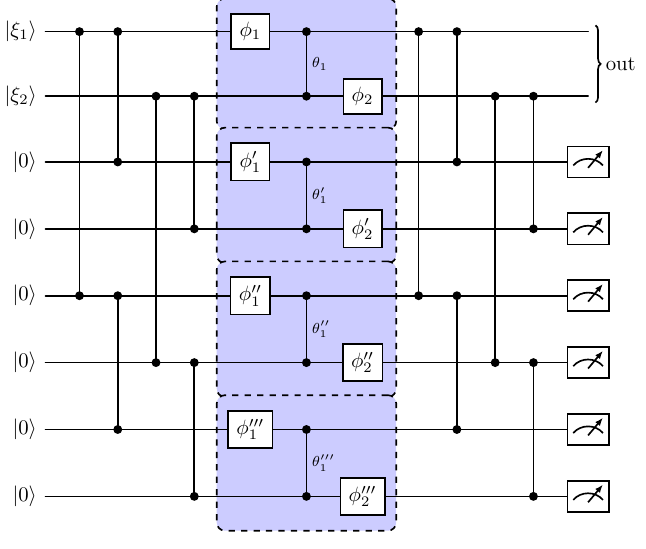}
		\end{subfigure}
		\caption{The circuit diagram depicts the unitary averaging with continuous-variables (UA) protocol for a two-mode system with $n=1$ (Fig.\ref{fig:2MODE1}), $n =2$ (Fig.\ref{fig:2MODE2}) and  $n =4$ (Fig.\ref{fig:2MODE4}). Fig.\ref{fig:2MODE1} corresponds to the $n=1$ case, where the UA protocol is not applied. Fig.\ref{fig:2MODE2} represents the $n=2$ scenario, which includes one additional replica of the unitary. Lastly, Fig.\ref{fig:2MODE4} illustrates the $n=4$ case, featuring three extra replicas of the unitary.
        Each state $\ket{\xi_{i}}$ represents a single-mode squeezed vacuum state with a squeezing factor $r_{i}$. The operator $\phi_{j}$ denotes the phase rotation applied to the $j$th mode.
In circuit diagrams, a solid line connecting two dots represents a 50:50 beamsplitter used in encoding and decoding transformations. Conversely, a solid line linking two dots ($\theta_{j}$) within a dashed box indicates a beamsplitter with adjustable transmissivity. $\phi_{j}'$ and $\theta_{j}'$ represents $\phi_{j}$ and $\theta_{j}$ with independent Gaussian phase noise, characterized by a mean of zero and a standard deviation of $\sigma$.
The dashed box highlights the error unitary, where the transformations on the left and right sides of the box correspond to the encoding and decoding processes, respectively. Vacuum detection serves as the measurement operator.
\label{CVUA2}}
\end{figure*}

\begin{figure*}[!htb]
	\centering
  \begin{subfigure}{\columnwidth}
  \caption{\label{fig:Fid2mode0.50.7}}
			\includegraphics[width=0.85\columnwidth]{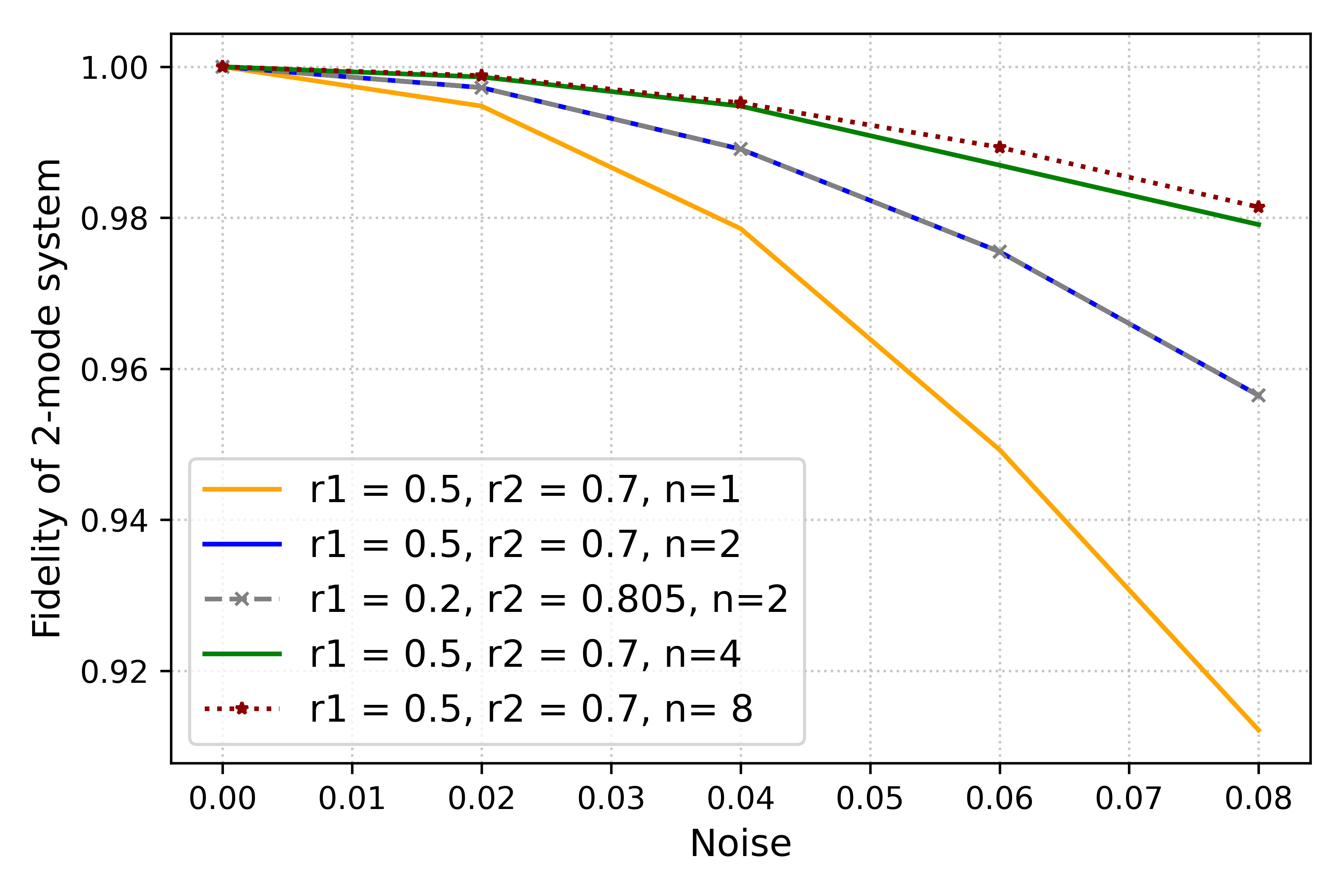}
			
		\end{subfigure}
  \begin{subfigure}{\columnwidth}
  \caption{\label{fig:2modeProbability}}
			\includegraphics[width=0.85\columnwidth]{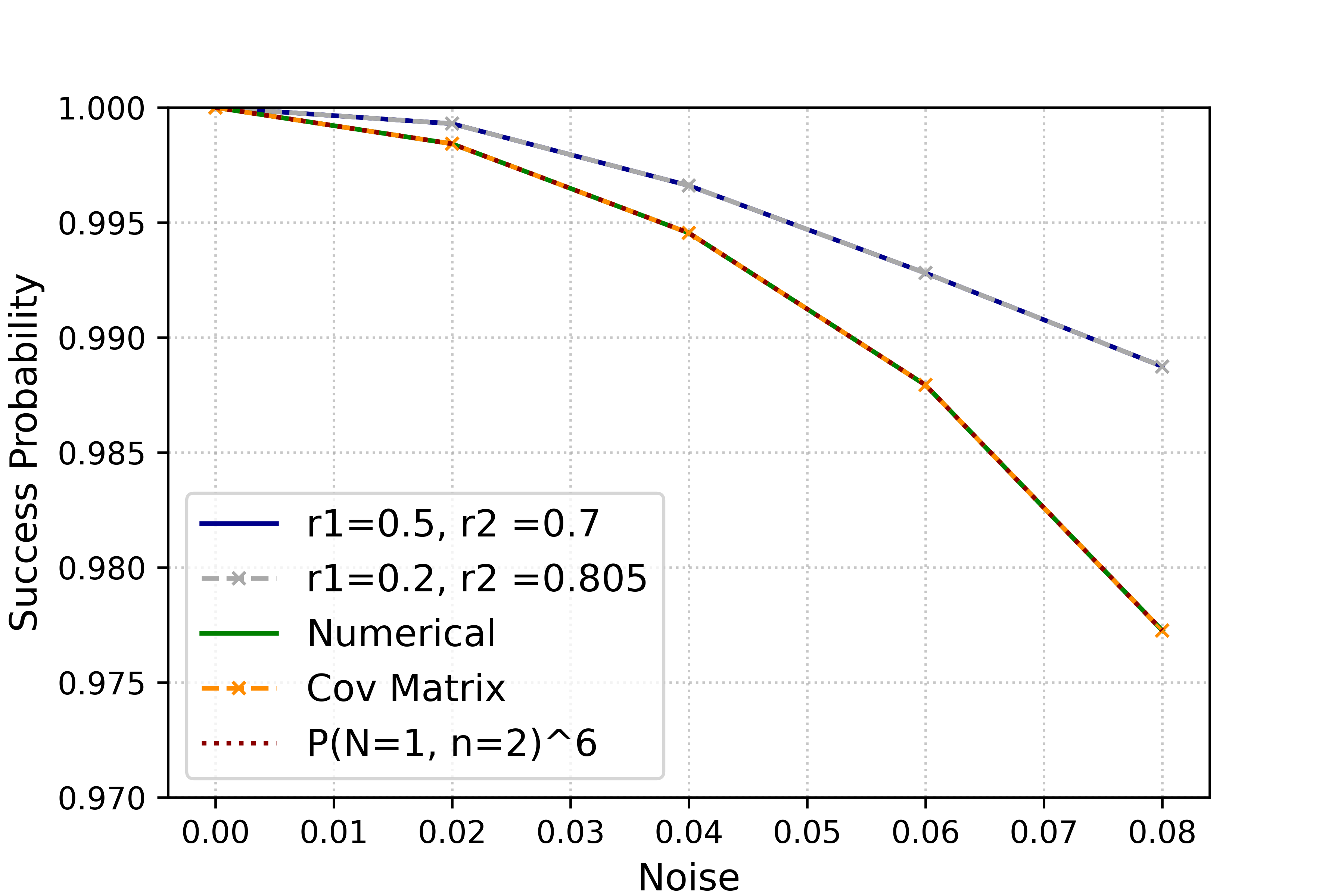}
		\end{subfigure}
		\caption{
        Fidelity and Success probability of two-mode system. 
       Fig.\ref{fig:Fid2mode0.50.7} corresponds to an average photon number of approximately $0.84$. In solid lines, the squeezing factors are  $r_{1} = 0.5, \: r_{2} = 0.7$, while in dotted green lines, they are $r_{1} = 0.2, \: r_{2} = 0.805$. The fidelity improvement is evident, though it begins to saturate from $n=8$ onwards. The most significant improvement is observed in the $n=2$ case. The orange line, blue solid line, green solid line, and purple dotted line represent the UA protocol for $n=1$, $n=2$, $n=4$, $n =8$, respectively.
(\ref{fig:2modeProbability}): The solid green line represents the probability calculated using Eq.\eqref{Fockcal}, while the orange dashed line with cross marker corresponds to the result obtained from Eq.\eqref{decodeCov} using the average photon number described in the text. The dotted red line represents the result derived from the power law formula \eqref{Probguru}. All three methods demonstrate strong agreement in their success probabilities.
In Fig.\ref{fig:2modeProbability}, the squeezing factor is set to $r_{1} = r_{2} = r = 0.7$. By maintaining the same average photon number (approximately $0.84$), the success probability of the UA protocol remains unaffected by variations in the squeezing factor across modes. The solid blue line represents $r_{1} = 0.5, \: r_{2} = 0.7$, while the dashed gray line shows $r_{1} = 0.2, \: r_{2} = 0.805$. All the success probability plots are based on UA with $n=2$.
  \label{fig:CVUA2modeplots}}
\end{figure*}

\begin{figure*}[!htb]
	\centering
		\begin{subfigure}{\columnwidth}
             \caption{ \label{fig:3MODE}}
			\includegraphics[width=1.0\columnwidth]{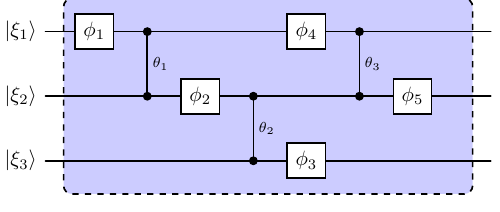}
			
		\end{subfigure}
  		\begin{subfigure}{\columnwidth}
              \caption{ \label{fig:4MODE}}
			\includegraphics[width=\columnwidth]{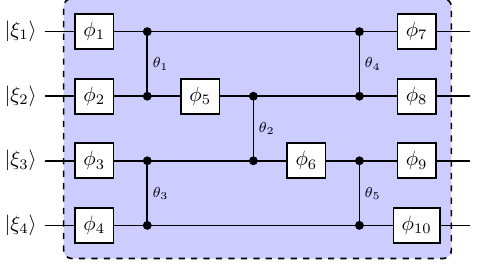}
		\end{subfigure}
        \begin{subfigure}{\columnwidth}
  \caption{\label{fig:5MODE}}
			\includegraphics[width=1.0\columnwidth]{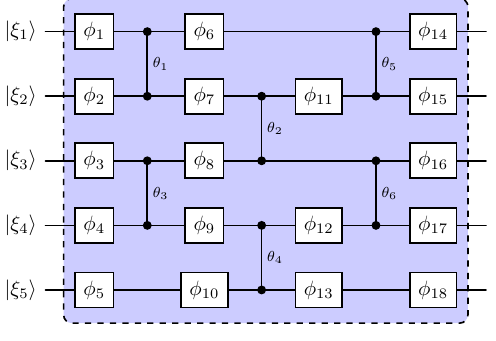}
		\end{subfigure}
        \begin{subfigure}{\columnwidth}
  \caption{\label{fig:NMODECVUA}}
			\includegraphics[width=0.65\columnwidth]{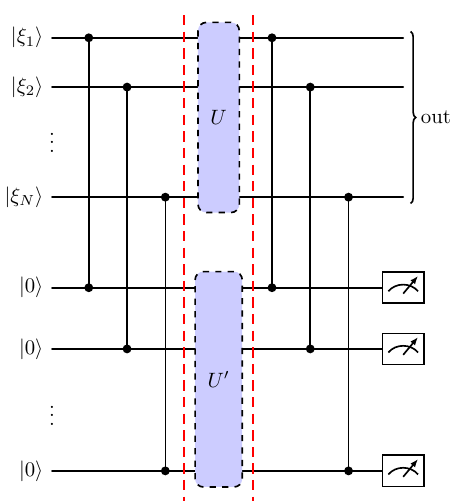}
		\end{subfigure}
		\caption{Multiport interferometers and UA protocol.
         Each state, denoted as  $\ket{\xi_{i}}$, represents a single-mode squeezed vacuum state characterized by a squeezing factor $r_{i}$. The operator $\phi_{j}$ corresponds to a phase rotation applied to the $j$-th mode. In circuit diagrams, a solid line connecting two dots represents a 50:50 beamsplitter used in encoding and decoding transformations. Conversely, a solid line linking two dots within a dashed box indicates a beamsplitter with adjustable transmissivity. The dashed box highlights a general interferometer featuring noisy gates. Specifically: Figure (\ref{fig:3MODE}) illustrates a general interferometer for a 3-mode system.
Figure (\ref{fig:4MODE}) illustrates a general interferometer for a 4-mode system. Figure (\ref{fig:5MODE}) illustrates a general interferometer for 5-mode system. Lastly, Figure (\ref{fig:NMODECVUA}) depicts the general N-mode UA protocol for the case $n=2$. In this scenario, the transformations on the left and right sides of the dashed boxes represent the encoding and decoding processes, respectively. Vacuum detection is employed as the measurement operator.
  \label{fig:NMODECIRCUIT}}
\end{figure*}

We will initially present the simplest case, a two-mode system.
Consider a two mode system where two single-mode squeezed vacuum input states undergo an arbitrary linear optical transformation, as depicted in the Fig.\ref{fig:2MODE1}. The input state is then
\begin{align}
   \ket{ \psi_{\text{In}}} &=  \ket{\xi_{1}} \ket{\xi_{2}} \nonumber\\
                           & = e^{\frac{r_{1}}{2}( e^{-i \phi_{1}} \hat{a}^{2} -  e^{i \phi_{1}} \hat{a}^{\dag 2})}e^{\frac{r_{2}}{2}( e^{-i \phi_{2}} \hat{a}^{2} - e^{i \phi_{2}} \hat{a}^{\dag 2})} \ket{0}\ket{0}
\end{align}
$r_{j}$ is the squeezing factor on mode $j$. $\hat{a}$ and $\hat{a}^{\dag}$ are annihilation and creation operator respectively. The circuit in the  Fig.\ref{fig:2MODE1} inside the purple box applies an arbitrary linear optical unitary \cite{clements2016optimal} which can be written as $\hat U =\hat{\phi_{1}} \hat{B}_{12}(\theta_{1}) \hat{\phi_{2}} $, which consisting of phase shifters (white boxes) and a beamsplitter (solid lines linking two dots) with the indicated parameters.
The beamsplitter transformation acting on modes $i$ and $j$ is in general defined as,
\begin{align}
    \hat{B}_{ij}(\theta) & = \exp{\theta (\hat{a}^{\dag}_{i}\hat{a}_{j} - \hat{a}_{i}\hat{a}^{\dag}_{j} )}
\end{align}

The beam splitter transmissivity is determined by the angles $\theta$, $\theta_{k}^{'}$, $\theta_{k}^{''}$, and $\theta_{k}^{'''}$ in Fig.\ref{CVUA2}, $k = 1, ..., N$. Specifically, the transmissivity is given by $\eta = \cos^{2}\theta_{k}$, $\eta = \cos^{2}\theta_{k}^{''}$, and $\eta = \cos^{2}\theta_{k}^{'''}$, respectively.

The phase rotation operator with angle $\hat{\phi}_j$ acting on mode $j$ is defined
\begin{align}
   \hat{ \phi}_{j} = e^{- i \phi_j \hat{a}^{\dag} \hat{a}} \label{ROOP}
\end{align}

In Fig.\ref{CVUA2}, $\hat{\phi}^{'}_{j}$, $\hat{\phi}^{''}_{j}$ and $\hat{\phi}^{'''}_{j}$ also defined as in Eq.\ref{ROOP}. 
Beamsplitters and phase shifters are defined in more detail in Appendix \ref{appendixA0}.
 The final state after applying the unitary is given by $\ket{\psi}_{\text{Out}} = \hat{\phi_{1}} \hat{B}(\theta_{1}) \hat{\phi_{2}} \ket{\psi_{\text{In}}}$, where step-by-step calculations are provided in the Appendix \ref{appendixA}. We now introduce noise into our circuit by allowing each of the parameters to fluctuate stochastically. We consider each component in $\hat{U}$ has random equal and independent Gaussian noise.  That is all the $\phi_j$'s and $\theta$'s fluctuate around there respective means. The variance of these Gaussian fluctuations is labelled as ``noise" in the figures. In Fig.\ref{fig:Fid2mode0.50.7} the orange dotted line shows the fidelity of the two-mode system with noise.

By applying the UA protocol once ($n=2$) which involves adding one additional replica of the unitary, the input state:
\begin{align}
     \ket{ \psi_{\text{In}}}  &=  \ket{\xi_{1}} \ket{\xi_{2}} \ket{0}\ket{0} \label{inputFock}
\end{align}
is first encoded using 50:50 beamsplitters according to
\begin{equation}
    \hat{B}_{\text{Encode}} = \hat{B}_{13} \times \hat{B}_{24}
\end{equation}
before being transformed with $\hat{U}$ and $\hat{U}^{'}$ as ideally identical copies of one another, with independent stochastic noise thus giving
\begin{align}
    \hat{U}_{UA} =& \hat{U} \otimes \hat{U}^{'} \nonumber\\
    =& \left(\hat{\phi_{2}}  \hat{\phi_{2}}' \hat{B}_{12}(\theta) \hat{B}_{34}(\theta') \hat{\phi_{1}}  \hat{\phi_{1}'}\right)
\end{align}
before being decoded, again with 50:50 beamsplitters
\begin{equation}
    \hat{B}_{\text{Decode}}  = (\hat{B}_{13} \times \hat{B}_{24})^{\dag}.
\end{equation}
After heralding vacuum on the last two-modes, our final output state is
\begin{equation}
   \ket{\psi}_{\text{Out}} =  \mathcal{N}_{UA}~_3\langle 0|_4\langle 0|  \hat{B}_{\text{Encode}}  \hat{U}_{UA} \hat{B}_{\text{Decode}} \ket{ \psi_{\text{In}}}  \label{Fockcal}
\end{equation}
 where $\mathcal{N}_{UA}$ is the normalisation constant which is related to the probability of success.
 
Fig.\ref{fig:CVUA2modeplots} presents the fidelity (Eq.\ref{fidelity}) and success probability of two-mode system. A clear improvement in fidelity is observed with the application of UA. Without averaging, the fidelity stands at $92\%$, whereas UA boosts it to $98\%$ at noise variance of 0.08 in Fig.\ref{fig:Fid2mode0.50.7}. Furthermore, we also demonstrate that when the average photon number is held constant, the fidelity and success probability remain consistent across different squeezing levels. To maintain a constant fidelity and success probability with respect to the average photon number, it is necessary to apply a certain amount of squeezing in each mode. If any mode is left unsqueezed, the fidelity will no longer remain constant with the average photon number.
This is evident in Figures \ref{fig:Fid2mode0.50.7} and \ref{fig:2modeProbability}, both the blue (dashed and solid) show a fidelity and success probability corresponding to an average photon number of approximately 0.84. In solid blue line, the squeezing parameters are $r_{1} = 0.5, \: r_{2} = 0.7$, while in dashed blue line, they are $r_{1} = 0.2, \: r_{2} = 0.805$. 
The success probability exceeds $97.5\%$ at noise is equal to 0.08 under practical noise conditions and squeezing level as illustrated in Fig. \ref{fig:2modeProbability}.

\begin{figure*}[!htb]
	\centering
		\begin{subfigure}{\columnwidth}
             \caption{ \label{fig:3modeplot}}
			\includegraphics[width=\columnwidth]{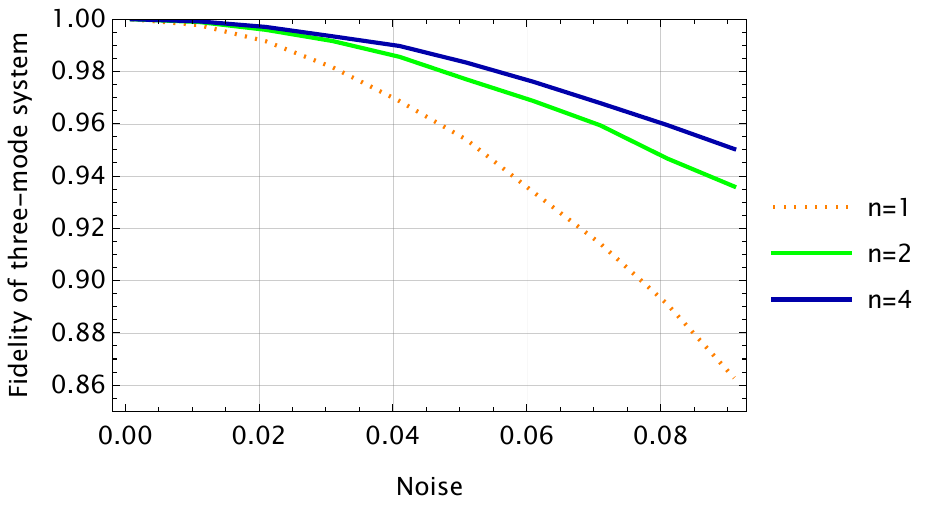}
			
		\end{subfigure}
  		\begin{subfigure}{\columnwidth}
              \caption{ \label{fig:4modeplot}}
			\includegraphics[width=\columnwidth]{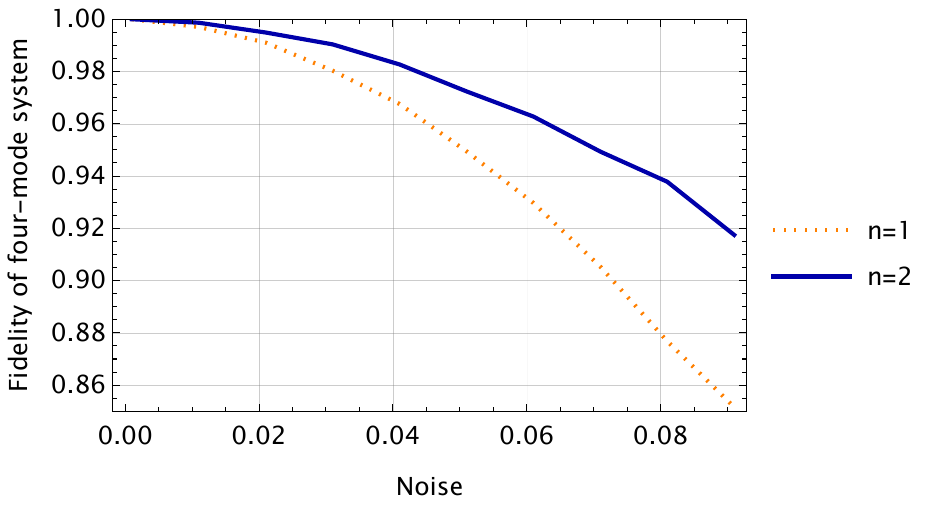}
			
		\end{subfigure}
		\begin{subfigure}{\columnwidth}
              \caption{ \label{fig:5modeplot}}
			\includegraphics[width=\columnwidth]{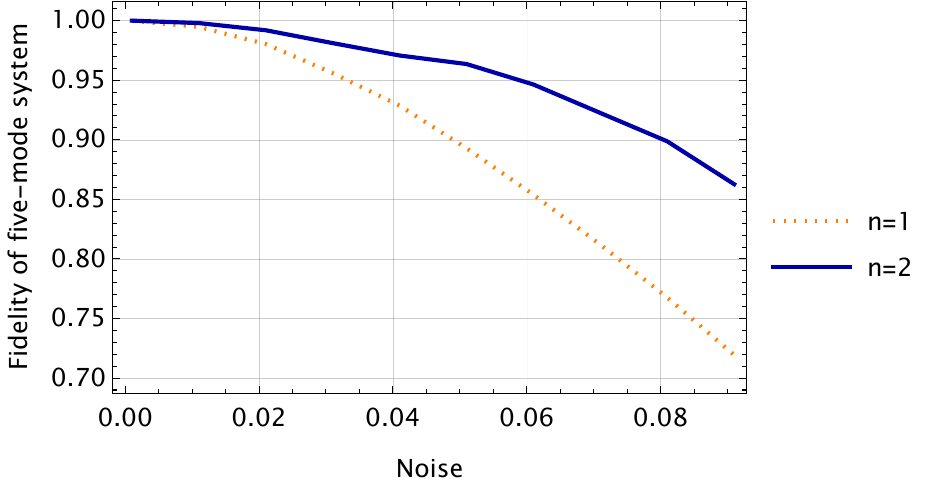}
			
		\end{subfigure}
		\begin{subfigure}{\columnwidth}
  \caption{ \label{10modesystem}}
			\includegraphics[width=0.85\columnwidth]{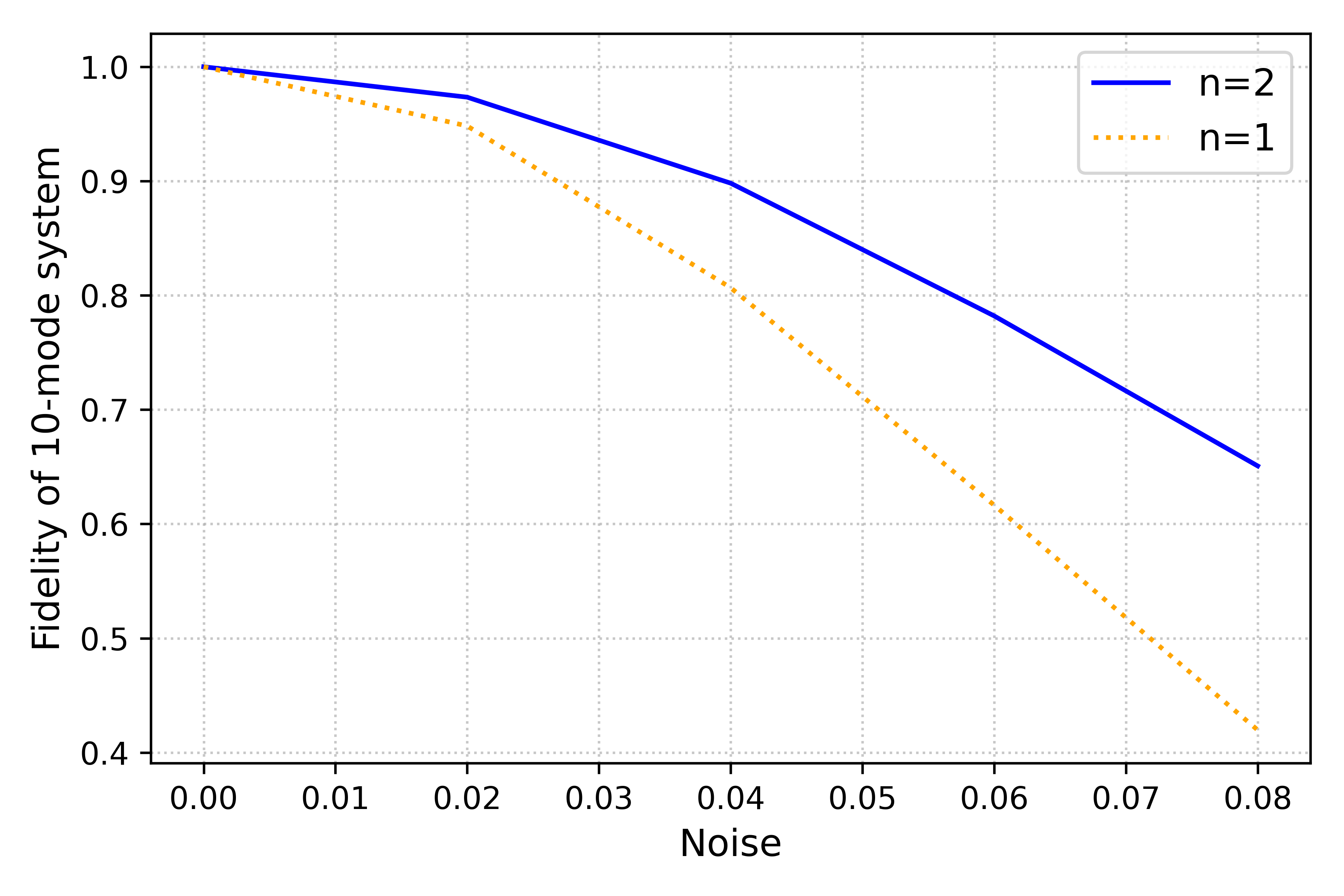}
			
		\end{subfigure}
		\caption{ Fidelity of 3,4, 5 and 10-mode systems with UA protocol in Figs. \ref{fig:3modeplot}, \ref{fig:4modeplot}, \ref{fig:5modeplot} and \ref{10modesystem} respectively. Here, $n$ represents the number of times the UA protocol is applied. In (\ref{fig:3modeplot}), the dotted orange line indicates the fidelity without the UA protocol. When the UA protocol is applied with $n=2$, the fidelity improves, as shown by the solid green line. The solid blue line corresponds to the fidelity when $n=4$. The squeezing factors are $r_{1} = 0.5, r_{2} = 0.6, r_{3} = 0.7$ for three-mode system. In \ref{fig:4modeplot} and \ref{fig:5modeplot}, the dotted orange line indicates the fidelity without the UA protocol. When the UA protocol is applied with $n=2$, the fidelity improves, as shown by the solid blue line. The squeezing factors are $r_{1} = 0.3, r_{2} = 0.4, r_{3} = 0.5, r_{4} = 0.6$ and $r_{1} = 0.3, r_{2} = 0.4, r_{3} = 0.5, r_{4} = 0.6,  r_{5} = 0.7$   for four-mode and five-mode system respectively.  (\ref{10modesystem}): The plot illustrates the fidelity of a 10-mode system as a function of noise with $r_{i} = 0.5, \: i = 1, \dots, 10$ on each modes. The dotted orange line represents the fidelity without the UA protocol, while the solid blue line demonstrates the enhanced fidelity achieved with the UA protocol in place. }
    \label{fig:345modesCVUA}
		\end{figure*}

\begin{figure*}[!htb]
	\centering
  \begin{subfigure}{\columnwidth}
  \caption{\label{10modepowerFid}}
			\includegraphics[width=0.85\columnwidth]{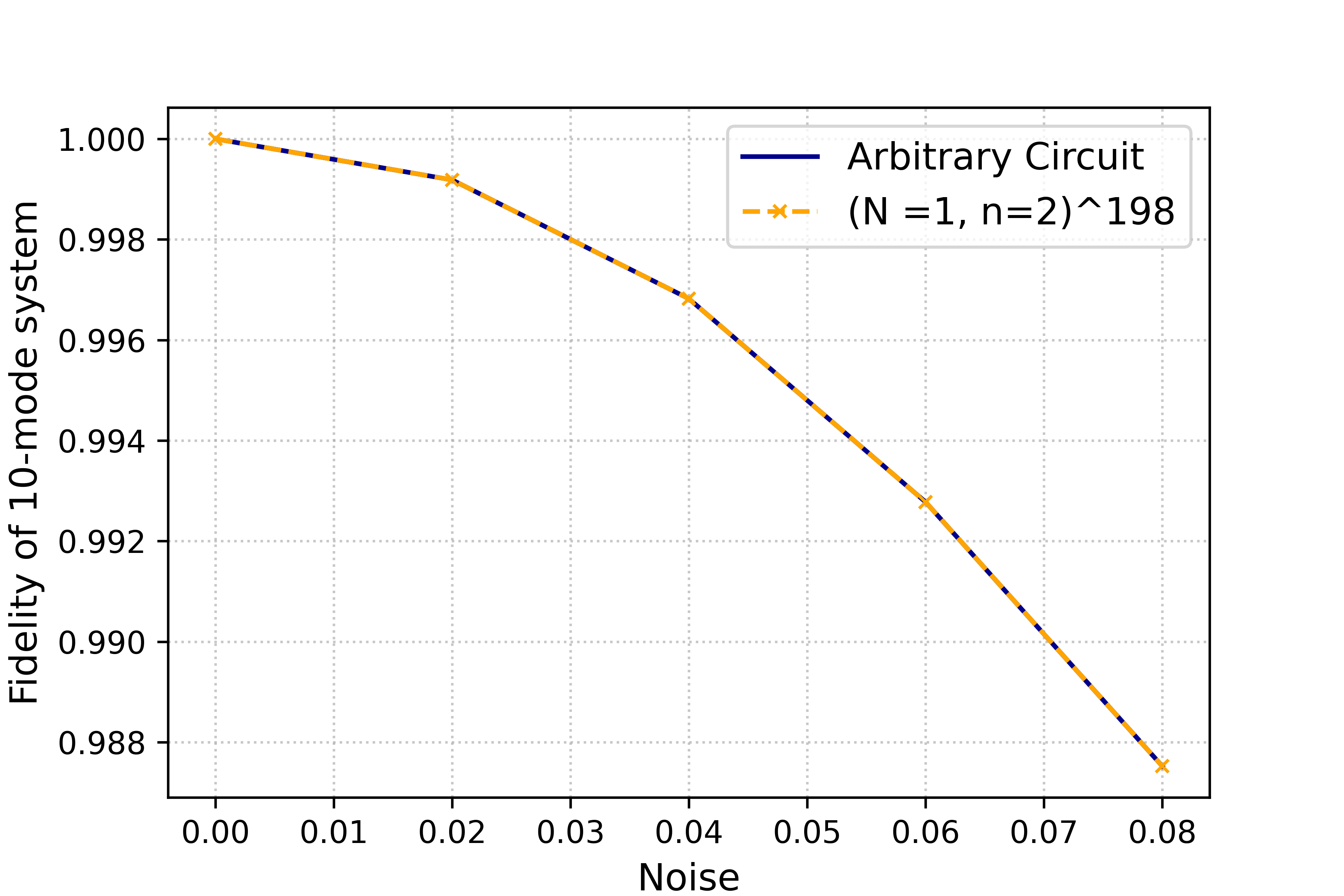}
			
		\end{subfigure}
  \begin{subfigure}{\columnwidth}
  \caption{\label{10modepowerProb}}
			\includegraphics[width=0.85\columnwidth]{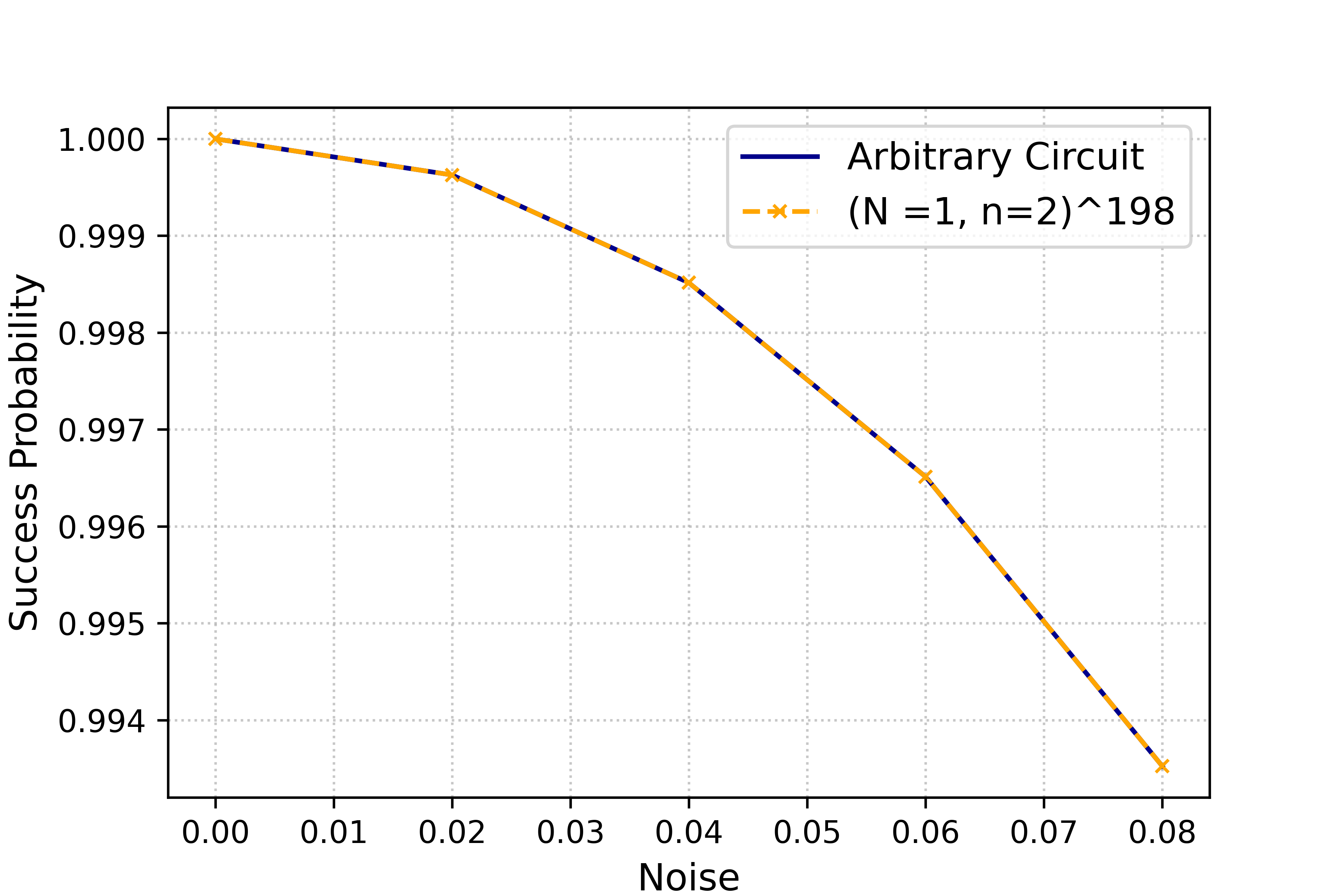}
		\end{subfigure}
		\caption{
 UA protocol with 10-mode system.
        Fig. (\ref{10modepowerFid}) illustrates the agreement with the power law \eqref{Fidpower}. The solid blue line depicts the fidelity achieved using the UA protocol, while the dashed orange line corresponds to the fidelity predicted by the power law formula. Fig. (\ref{10modepowerProb}) illustrates the agreement with the power law \eqref{Probpower}. The solid blue line depicts the probability achieved using the UA protocol, while the dashed orange line  corresponds to the probability predicted by the power law formula.
        Here, $N$ denotes the number of input modes, and $n$ represents the number of UA protocol repetitions (i.e., the number of times the protocol is applied). The squeezing parameter for the first 10 modes is set to 0.1, meaning $r_{1} = r_{2} = ...= r_{10} = 0.1$. Since the numerical calculation of \ref{10modesystem} and \ref{10modepowerFid} is highly time-consuming, we have presented the plots using a significantly reduced noise scale.
  \label{fig:CVUA10}}
\end{figure*}


\begin{figure*}[!htb]
	\centering
  \begin{subfigure}{\columnwidth}
  \caption{\label{fig:4838noisy}}
			\includegraphics[width=0.85\columnwidth]{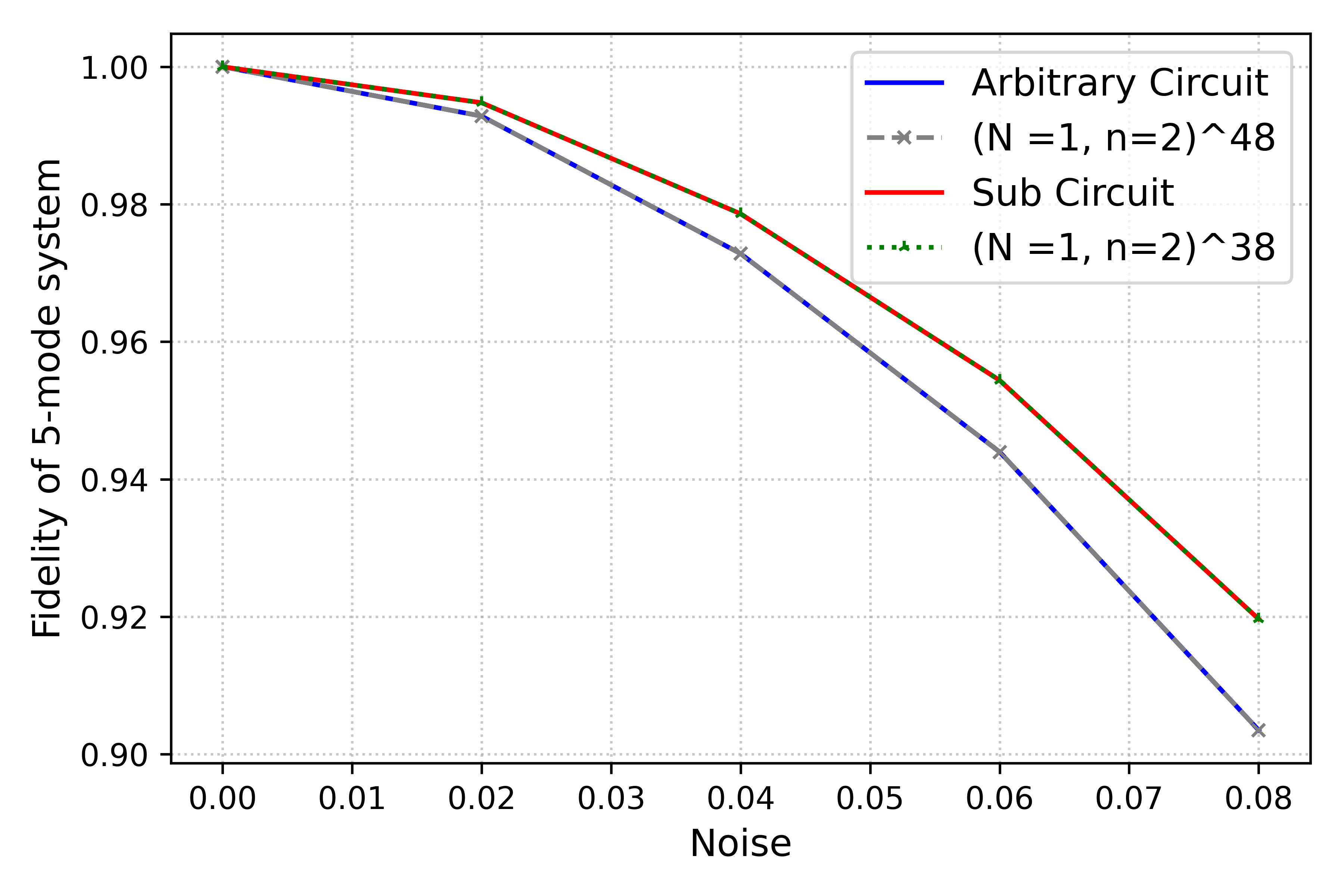}
			
		\end{subfigure}
  \begin{subfigure}{\columnwidth}
  \caption{\label{fig:prob4838}}
			\includegraphics[width=0.85\columnwidth]{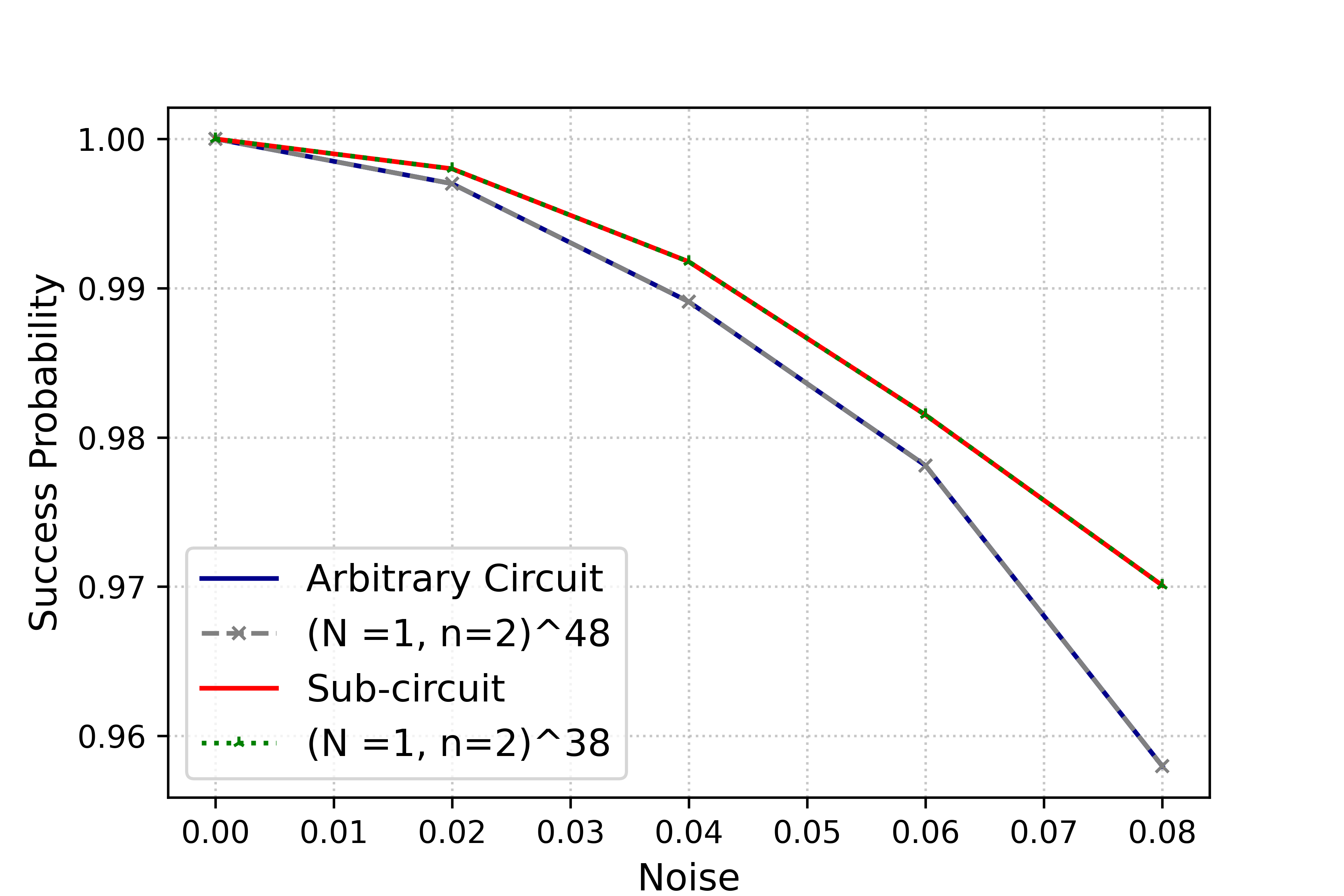}
		\end{subfigure}
		\caption{ Validation of the power law expression for fidelity \eqref{Fidpower} and probability \eqref{Probpower} . (\ref{fig:4838noisy}) and  (\ref{fig:prob4838}) depict the fidelity and probability of a five-mode system with an arbitrary circuit and sub-circuit, where each mode has a squeezing factor of $r_1 = r_2 = r_3 = r_4 = r_5 = 0.5$. The solid blue line corresponds to the fidelity and probability from the numerical calculation, while the dashed gray line represents the result from \eqref{Fidpower} and \eqref{Probpower} with $N^2 - 1 = 24$, respectively. Similarly, solid red lines represents results from numerical calculation while the dotted green line shows from power law formula with $N^2 - 1 = 19$.
  \label{fig:noisyFid}}
\end{figure*}

\begin{figure*}[!htb]
	\centering
		\begin{subfigure}{\columnwidth}
             \caption{ \label{fig:50mode}}
			\includegraphics[width=\columnwidth]{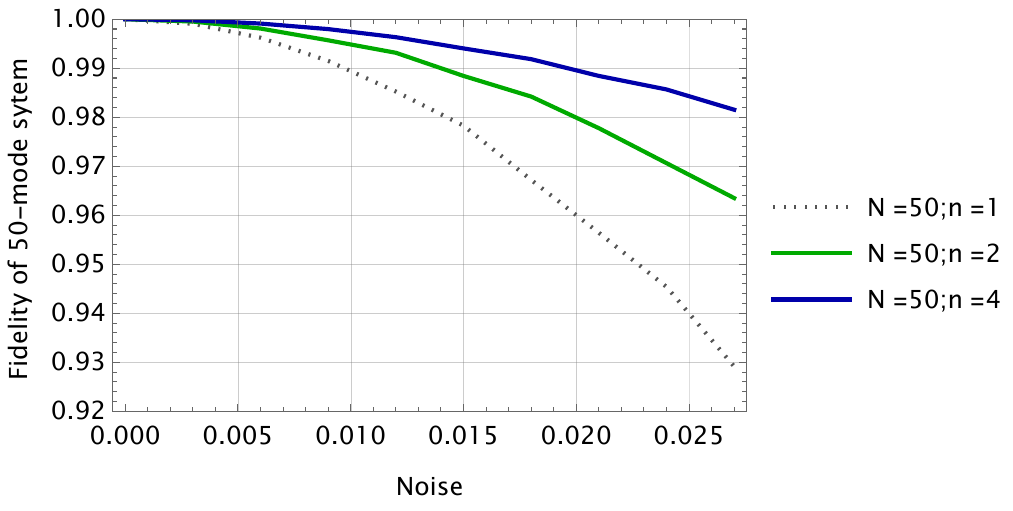}
			
		\end{subfigure}
  		\begin{subfigure}{\columnwidth}
              \caption{ \label{fig:100mode}}
			\includegraphics[width=\columnwidth]{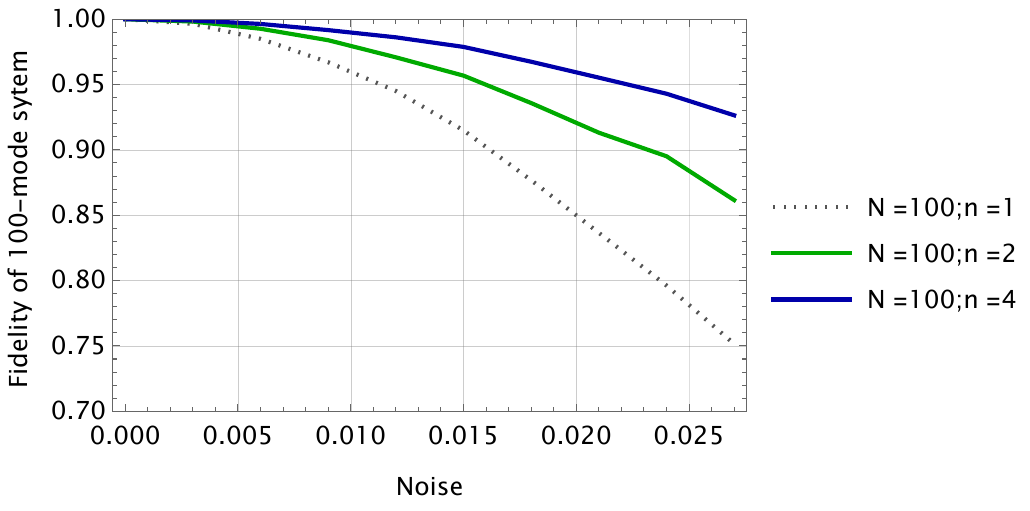}
			
		\end{subfigure}
		\begin{subfigure}{\columnwidth}
              \caption{ \label{fig:150mode}}
			\includegraphics[width=\columnwidth]{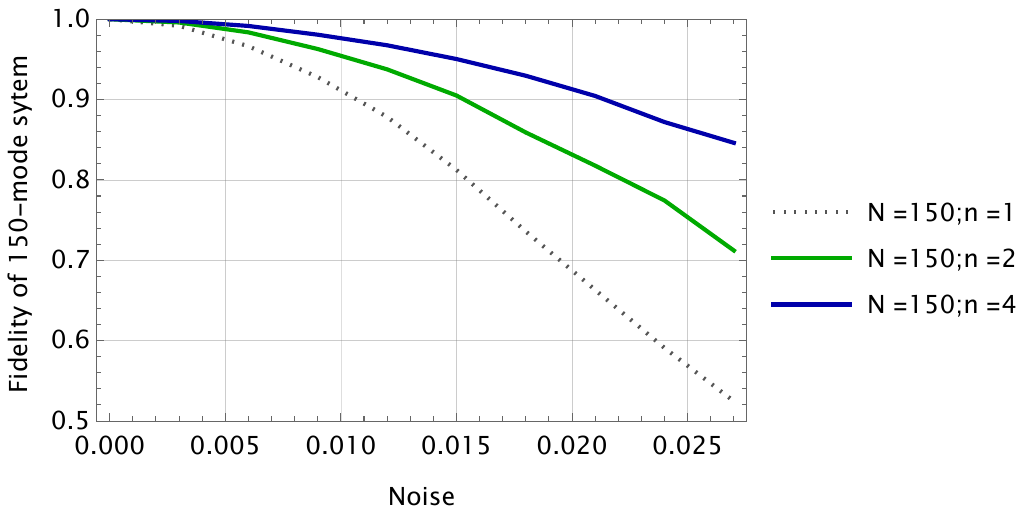}
			
		\end{subfigure}
		\begin{subfigure}{\columnwidth}
  \caption{ \label{fig:216mode}}
			\includegraphics[width=\columnwidth]{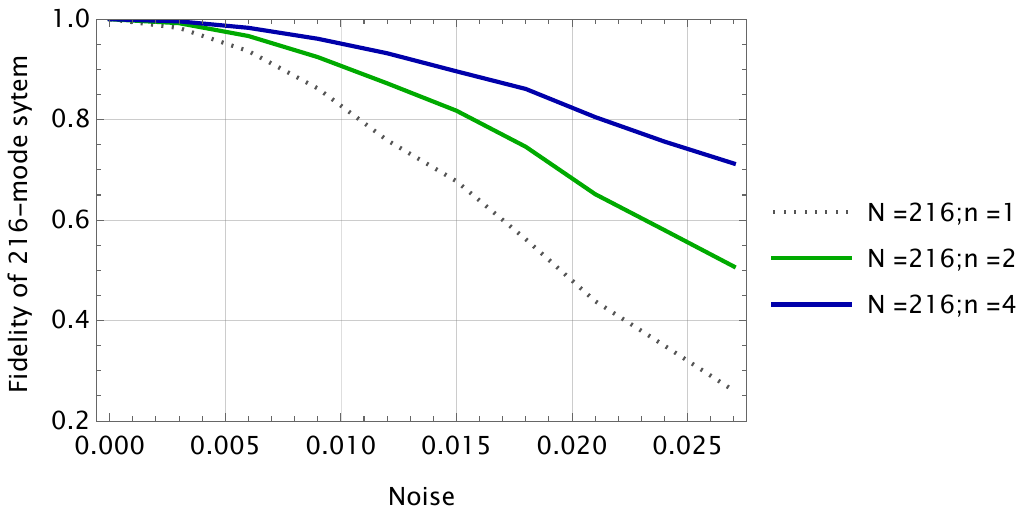}
			
		\end{subfigure}
		\caption{ The fidelity of the higher-mode systems using a power-law model \eqref{Fidpower} is illustrated in Figures \ref{fig:50mode}, \ref{fig:100mode}, \ref{fig:150mode}, and \ref{fig:216mode}. Here, $N$ denotes the number of modes, and $n$ represents the number of times the UA protocol is applied with squeezing $r_{i} = 0.1$ on each $i $ mode. The dotted gray line indicates the fidelity without the UA protocol. When the UA protocol is applied with $n=2$, the fidelity improves, as shown by the solid green line. The solid blue line corresponds to the fidelity when $n=4$.
\label{fig:extrapolated plots}}
		\end{figure*}

\begin{figure*}[!htb]
	\centering
		\begin{subfigure}{\columnwidth}
             \caption{ \label{fig:50modeprob}}
			\includegraphics[width=\columnwidth]{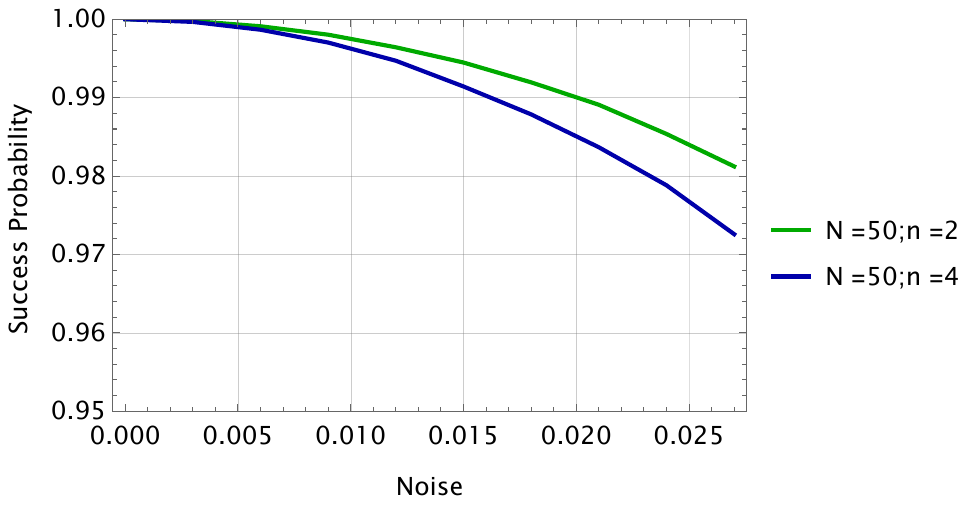}
			
		\end{subfigure}
  		\begin{subfigure}{\columnwidth}
              \caption{ \label{fig:100modeprob}}
			\includegraphics[width=\columnwidth]{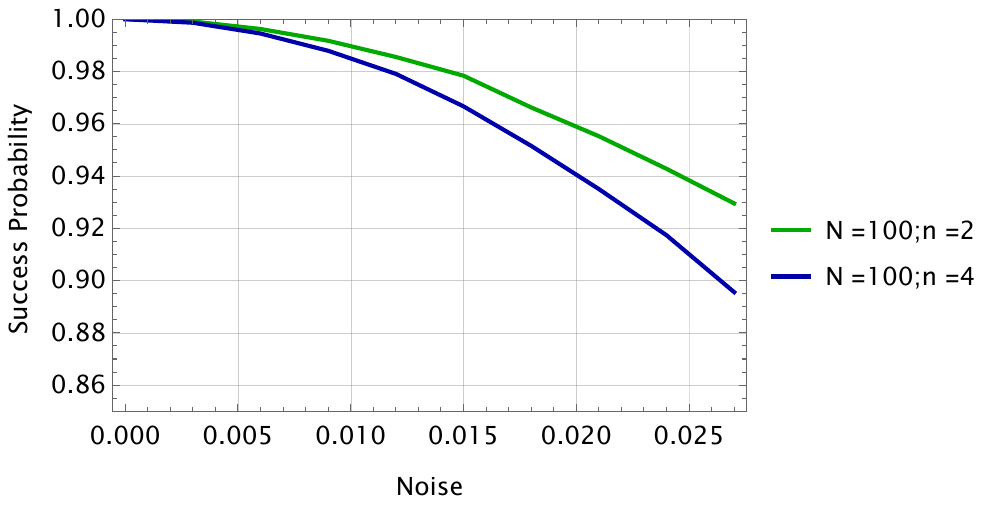}
			
		\end{subfigure}
		\begin{subfigure}{\columnwidth}
              \caption{ \label{fig:150modeprob}}
			\includegraphics[width=\columnwidth]{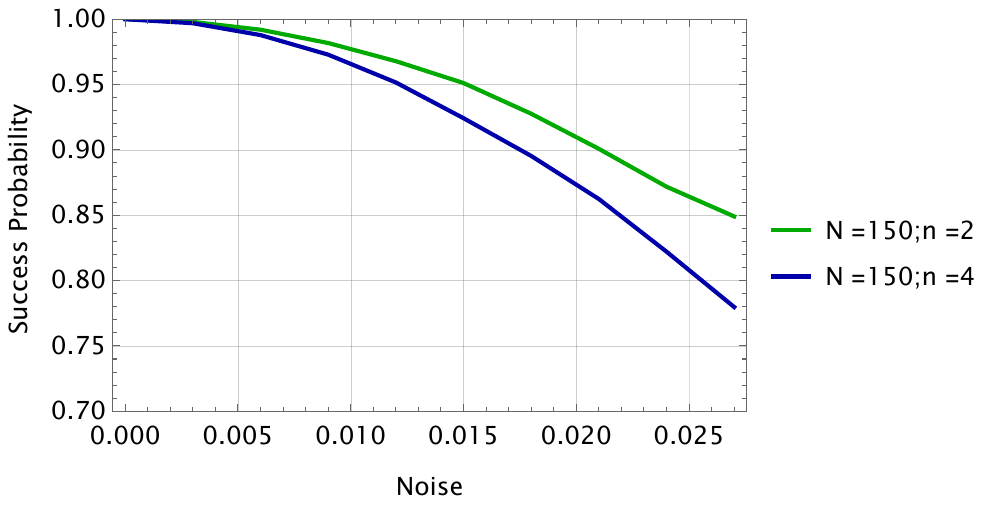}
			
		\end{subfigure}
		\begin{subfigure}{\columnwidth}
  \caption{ \label{fig:216modeprob}}
			\includegraphics[width=\columnwidth]{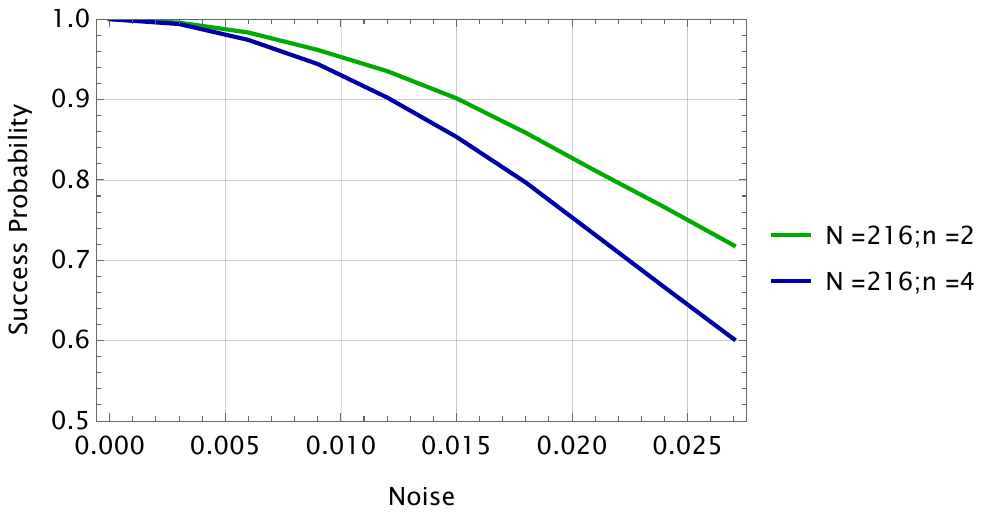}
			
		\end{subfigure}
		\caption{ The probability of the higher-mode systems using a power-law model \eqref{Probpower} is illustrated in Figures \ref{fig:50modeprob}, \ref{fig:100modeprob}, \ref{fig:150modeprob}, and \ref{fig:216modeprob}. Here, $N$ denotes the number of modes, and $n$ represents the number of times the UA protocol is applied with squeezing $r_{i} = 0.1$ on each $i $ mode.  The solid green line represents the success probability with $n =2$. The solid blue line corresponds to the probability when $n=4$.
\label{fig:extrapolated plotsprob}}
		\end{figure*} 
        
			

\begin{figure}[!htb]
	\centering
		 \includegraphics[width=0.85\linewidth]{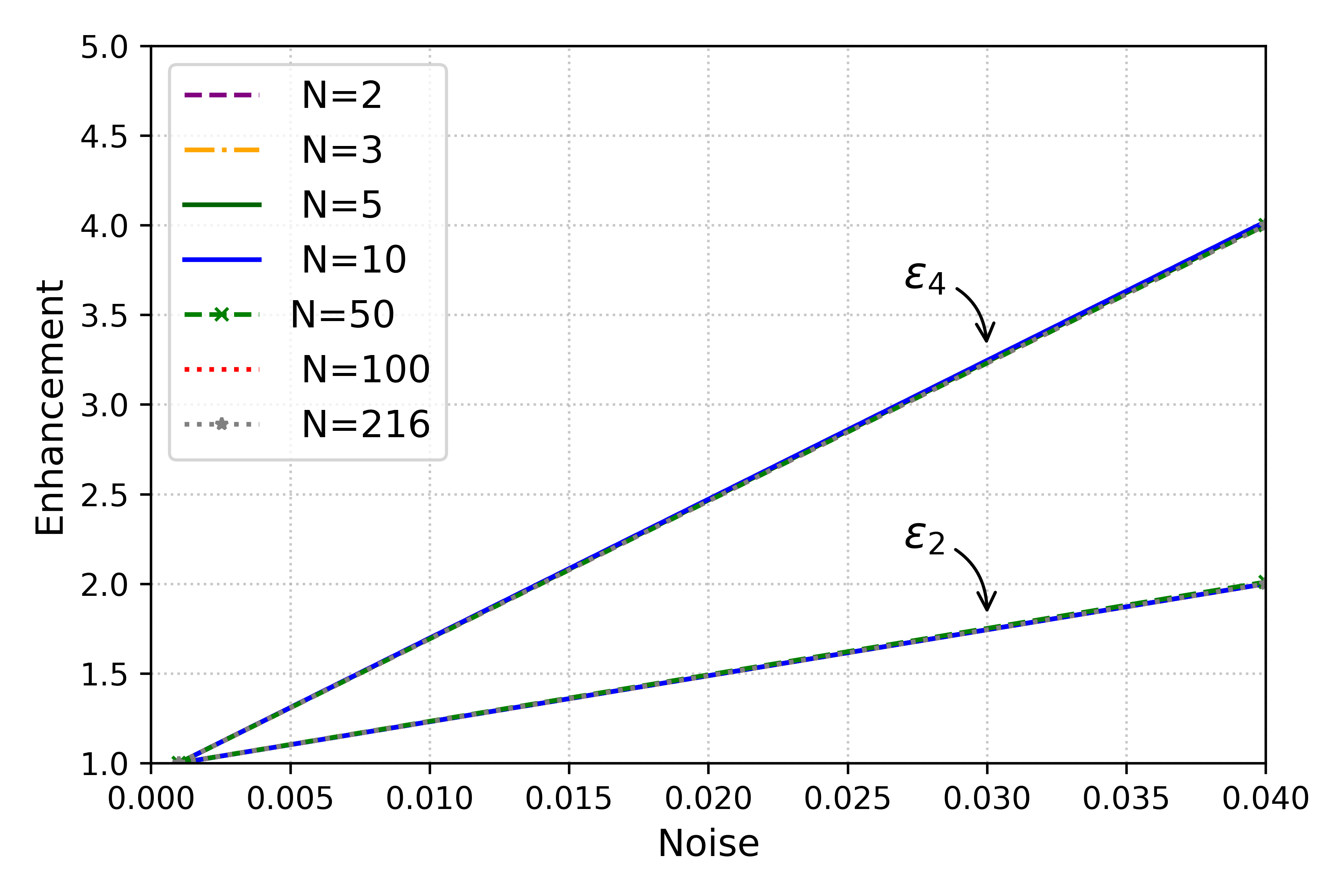}
		\caption{ Enhancement (Eq. \ref{EN}) for a multimode system. It illustrates that the enhancement remains identical for any multimode system in the presence of noise. Here, squeezing is $r_{i} = 0.1$ for ith mode. Enhancement ranges from 1 to 2 for UA with $n = 2$ (annotated as $\epsilon_{2}$), and from 1 to 4 for UA with $n = 4$ (annotated as $\epsilon_{4}$).
  \label{fig:enhance}}
\end{figure}

\subsection{Scaling and Approximation}

We evaluate the success probability numerically, noting that the Fock space grows exponentially with the number of particles and modes. For a system with larger particles and higher available modes, the number of possible states grows combinatorially, rendering computations intractable for larger systems. Additionally, storing and manipulating extensive Fock space representations demands substantial memory. 
To address these challenges and enable faster computations, we employ alternative representations such as symplectic transformation analysis to efficiently quantify the output state.
The output state is only approximately Gaussian for small noise, with this approximation becoming exact as the noise vanishes. This analysis relies on the assumption of Gaussianity. In \cite{swain2024improving}, we explored this approximation by directly calculating the fourth-order moments and comparing them to the predicted values for a Gaussian distribution. In Appendix \ref{appendixB}, we present a clear and step-by-step algorithm  for a two-mode system that applies the UA protocol, which can also be extended to multi-mode systems.

Now we have VDecode, the output covariance matrix (from Appendix \ref{appendixB}), after encoding and decoding steps for two-mode system. One might question how vacuum measurement can be performed, given that vacuum measurement corresponds to a Fock measurement. Interestingly, we checked that performing heterodyne detection on a Gaussian state with zero mean is equivalent to vacuum detection (more details found in Appendix.\ref{vacdet}).
To measure the vacuum modes at the end and achieve our reduced noise output state, we perform a heterodyne measurement on zero as the vacuum detection on VDecode matrix.

Let, \begin{align}
   \text{VDecode} &= \begin{pmatrix}
       A & C \\
       C & B
   \end{pmatrix}   \label{decodeCov}
\end{align}
By heterodyning on zero the last mode, the first $N$ modes are still in a Gaussian state, and the output covariance matrix is given by

\begin{align}
    V_{\text{Out}} = A - C (B + \mathbf{I})^{-1} C^{T} \label{COVGuru}
\end{align}
where $\mathbf{I}$ is the $2 \times 2$ Identity matrix.

Consider two $N$-mode Gaussian states, $\rho_{0}$ and $\rho_{1}$, where $\rho_{0}$ is a pure state and $\rho_{1}$ is a mixed state. The fidelity between these states is given by 
\begin{align}
    F = \frac{2^{N}}{\sqrt{\text{det}(V_{0} + V_{1}})}  \label{fidelity}
\end{align}

where $V_{0}$ and $V_{1}$ denote the covariance matrices of the pure and mixed states, respectively, both assumed to have zero mean \cite{weedbrook2012gaussian, spedalieri2012limit}.
Determining the probability of a multi-mode system can be challenging, especially since methods like the Hafnian tend to overestimate results for lower-mode systems such as two or three modes \cite{quesada2018gaussian, kruse2019detailed}. To address this, we have used an alternative approach to calculate the success probability directly from the covariance matrix by tracing out the vacuum modes after the decoding step.
If the vacuum covariance matrix closely resembles a thermal state, we can estimate the success probability from the product of the average photon numbers. For instance, in the case of a UA with $n=2$ (Fig.\ref{fig:2MODE2}), the two vacuum modes form a $4 \times 4$ covariance matrix with four variances. The average photon numbers $( \mathbb{\bar{n}}_{1})$ and $( \mathbb{\bar{n}}_{2})$ are,
\begin{align}
    \mathbb{\bar{n}}_{1}  & = \sum_{i} P_{1}(i) \times i \\
    & = P_{1}(0) \times 0 + P_{1}(1) \times 1 + P_{1}(2) \times 2 +...  \nonumber\\
    \text{if} \: P_{1}(N) << 1 &  \: \text{for} \: N \geq 2  \nonumber\\
    \mathbb{\bar{n}}_{1}  & \approx P_{1}(1) \approx 1- P_{1}(0)  \\
    P_{1}(0) & \approx 1 - \mathbb{\bar{n}}_{1}
\end{align}

Similarly,
\begin{align}
    \mathbb{\bar{n}}_{2}  & \approx 1- P_{2}(0)  \\
    P_{2}(0) & \approx 1 - \mathbb{\bar{n}}_{2}
\end{align}
The average photon numbers are calculated as follows:
\begin{align}
     \mathbb{\bar{n}}_{1} & = 1/4 ((A_{11} +  A_{22}) -2 ) \\
     \mathbb{\bar{n}}_{2} & = 1/4 ((B_{33} +  B_{44}) -2 )
\end{align}
where $A_{ii}$ are variances of covariance matrix for two-mode system.
The success probability (P) is then given by:
\begin{align}
  \text{P} = (1- \mathbb{\bar{n}}_{1})(1- \mathbb{\bar{n}}_{2})  \label{Probguru}
\end{align}

We have calculated the success probability for modes $N=2, 3, 4, 5  \; \text{and} \; 10$ using Eq.\eqref{Probguru}. To verify the accuracy of this method, we compared the results with numerical calculations obtained from the Fock basis analysis based on Eq.\eqref{Fockcal} for the $n=2$ case, as illustrated in Fig.~\ref{fig:2modeProbability}. We have also calculated fidelity for modes $N=2, 3, 4, 5 \; \text{and} \;10$ from Eq.\eqref{COVGuru} using the multi-mode fidelity formula.
We have presented the complete numerical results for the two-mode system without any approximations in Fig.\ref{fig:CVUA2modeplots}. For the 3-mode, 4-mode, 5-mode and 10-mode systems, we provided both numerical fidelities and approximate success probabilities in Figs.\ref{fig:345modesCVUA} and \ref{fig:CVUA10}. Across all these figures, a clear trend emerges: fidelity consistently improves while maintaining a high success probability. A more detailed discussion of these results is provided in the discussion section.

\subsection{Large-scale System}

We have successfully generated results for systems with up to 10 modes, though we are limited by numerical constraints. The challenge now is how to simulate systems with more than 10 modes, as the computational complexity makes it numerically intractable to display results for larger systems. To address this, we are considering leveraging a relationship with the single-mode UA protocol, which is thoroughly explained in Section II. Since we already have an analytical expression for this protocol, it offers a potential pathway for extending our analysis to higher-dimensional systems.
Mathematically we make the ansatz:

\begin{align}
    F(N, n) & = F(N=1, n)^{x} \\
    P(N, n) & = P(N=1, n)^{x} 
\end{align}

Next, we attempted to identify the missing parameter, denoted as `x', by performing various checks against our numerical results. Through these checks, we determined that `x' corresponds to the number of noisy parameters, specifically ($2(N^{2} -1)$). We found that the formula aligns exactly with the numerical results when both squeezing and noise levels are low. This is particularly important, as it’s challenging to trust numerical results with high squeezing and noise due to the effects of truncation.
This is visually demonstrated in Figures (\ref{fig:CVUA10}, \ref{fig:noisyFid}). Interestingly, a recent study by Google Quantum has applied a similar power-law approach to define logical error rates \cite{acharya2024quantum}.

This approach, when combined with our previous analysis, offers a promising method to extend our results beyond the limitations of numerical tractability, suggesting a pathway forward for handling more complex, multi-mode systems.

The power law for fidelity (F) and probability (P) are, 
\begin{align}
    F(N, n) & = F(N=1, n)^{2(N^{2} -1)} \label{Fidpower}\\
    P(N, n) & = P(N=1, n)^{2(N^{2} -1)} \label{Probpower}
\end{align}

The expression $N^{2} -1$ represents the number of noisy parameters, where $N$ denotes the number of input modes. After verifying the solidness of the given relation, we utilised it to plot the fidelity of higher-mode systems for different values of $N$ in Fig.\ref{fig:extrapolated plots}. For a specific case $N=216$ (Fig.\ref{fig:216mode}), the fidelity initially stands at about $25 \%$, which we successfully improved to approximately $75 \%$ at the noise 0.03 with the application of UA. We have also plotted success probability of higher-mode systems in Fig.\ref{fig:extrapolated plotsprob}. 


\subsection{Enhancement}
We introduce a parameter called enhancement ($\epsilon$) to quantify the improvement in fidelity achieved through unitary averaging (UA). Mathematically, it is defined as-

\begin{align}
    \epsilon & = \frac{1-  F(N, n =1)}{1 - F(N, n)} \label{EN}
\end{align}

Here, $F(N, n=1)$ denotes the fidelity of an $N$-mode system without UA, while $F(N, n)$ represents the fidelity with UA. The fidelity can be obtained numerically or using the power-law relation given in Eq.~\eqref{Fidpower}.

We begin by evaluating enhancement with UA for $n = 2$, keeping the squeezing constant across all input modes. Interestingly, the enhancement remains consistent regardless of the number of input modes. We then examine the case where the squeezing varied across modes and find that the enhancement remained robust with the varying noise across any modes. However, in extreme scenarios—such as when one mode has no squeezing while the others are squeezed—the enhancement becomes inconsistent across modes with varying noise.

We extended our analysis to systems with $n = 4$, again observing consistent enhancement across modes with varying noise levels. These findings align with our earlier results for $n = 2$.

Extensive numerical evaluations for 2-, 3-, and 5-mode systems, supported by the results obtained using Eq.\eqref{Fidpower}, confirm that the enhancement $\epsilon$ remains consistent across different numbers of modes. This is illustrated in Fig.\ref{fig:enhance}, where enhancement ranges from 1 to 2 for UA with $n = 2$, and from 1 to 4 for UA with $n = 4$.

\section{Discussion and Conclusion}
The protocol requires only beamsplitters, phase shifters, and a standard click detector, along with the available input states. When working with squeezed states, an optical parametric amplifier is needed. All of these optical components are readily accessible in laboratories that conduct continuous-variable experiments.
The protocol can be implemented directly within a standard Gaussian Boson Sampling setup, without requiring any additional optical components beyond those already used in the circuit.

We will now discuss the results in more detail, starting with the two-mode system. Fig.\ref{fig:CVUA2modeplots} demonstrates a significant improvement in fidelity with a high success probability for the two-mode system. Notably, the fidelity remains above  $96 \%$ at the noise 0.1. For 3, 4, and 5-mode systems, the fidelity remains consistently high even in the presence of moderate squeezing and noise which is presented in Fig.(\ref{fig:345modesCVUA}). Similarly we have shown for  3-mode, 4-mode and 5-mode systems in Figs. \ref{fig:3modeplot}, \ref{fig:4modeplot} and \ref{fig:5modeplot} respectively. The success probability derived from Eq.\eqref{Probguru} is reliable when the vacuum covariance matrix closely resembles a thermal state. To illustrate this, we have plotted the fidelity and success probability of 3, 4, 5-mode and 10-mode systems in Fig.\ref{fig:345modesCVUA}. The results show that the success probability remains high, alongside high fidelity, confirming the solidness of our protocol. For the 10-mode system, where calculating fidelity is computationally demanding, we managed to demonstrate improved performance by considering a narrower noise range while still achieving a high success probability. Furthermore, our results validate the consistency of the power law formula with the numerical data obtained for the 10-mode system in Fig.\ref{fig:CVUA10}. For various modes, including 3, 4, and 5-mode systems, we have examined transformations with different noise parameters and demonstrated that our power law formula aligns closely with the numerical results presented in Fig.\ref{fig:noisyFid}. We observed that by maintaining the same average photon number, the success probability and fidelity of
the UA protocol remains unaffected by variations in the squeezing factor across modes.
For higher-mode systems, we have presented extrapolated results based on our power law model in Fig.\ref{fig:extrapolated plots}. We have also illustrated the enhancement for the multi-mode system in Fig.\ref{fig:enhance}. Interestingly, it remains consistent regardless of the number of modes with respect to the noise.
We also anticipate that the scheme is robust to loss as demonstrated in ref.\cite{swain2024improving}. Having established the robustness of the protocol, UA may have  practical applications across various domains, including communication and computation \cite{singh2023proof, huh2015boson, su2019conversion, aghaee2025scaling}.

\section{Acknowledgement}
SNS thanks Gerard Milburn, Nathan Walk and Michael Stefszky for helpful discussions.
SNS was supported by the Sydney Quantum Academy, Sydney, NSW, Australia.  
This work was partially supported by the Australian Research Council Centre of Excellence for Quantum Computation and Communication Technology (Project No.CE110001027).

\begin{appendix}
\section{Brief Review of the Phase-space Formalism}
\label{appendixA0}

As outlined in the main text, in order to analyse the noise reduction in Gaussian boson sampling circuits with unitary averaging, we employ the phase-space formalism \cite{BAC19,weedbrook2012gaussian, brask2021gaussian, ferraro2005gaussian, serafini2023quantum}. The following subsections are intended merely as a concise review of the relevant formalism to support the results presented in the main text.

\subsection{Mode and quadrature operators}
We consider bosonic systems composed of $N$ harmonic oscillators, each associated with annihilation and creation operators $\hat{a}_{j}$ and $\hat{a}^{\dag}_{j}$, where $j = 1, \dots, N$. The operators $\hat{a}_{j}$ will also be referred to as mode operators. Throughout, we set $\hbar = 2$ and adopt the canonical commutation relations.
\begin{align}
    [\hat{a}_{j}, \hat{a}_{k}] & = [\hat{a}^{\dag}_{j}, \hat{a}^{\dag}_{k}] = 0,\\
    [\hat{a}_{j}, \hat{a}^{\dag}_{k}] & = \delta_{jk}
\end{align}

We define Hermitian position- and momentum-like operators for each mode
\begin{align}
    \hat{x}_{j} & = (\hat{a}_{j} + \hat{a}^{\dag}_{j} ) \\
    \hat{p}_{j} & = i (  \hat{a}^{\dag}_{j} - \hat{a}_{j} )
\end{align}
with the canonical commutator
\begin{align}
    [\hat{x}_{j}, \hat{p}_{k}] & = 2i\delta_{jk}
\end{align}

Following the convention in quantum optics, we also refer to $\hat{x}_{j}$ and $\hat{p}_{j}$ as quadrature operators. For convenience, we collect the $2N$ quadrature operators corresponding to the $N$ modes into a single column vector, which we define as

\begin{align}
    \hat{r } = \begin{pmatrix}
        \hat{x}_{1} \\
        \hat{p}_{1} \\
        \vdots \\
        \hat{x}_{N} \\
        \hat{p}_{N}
    \end{pmatrix}
\end{align}
The canonical commutation relations can then be written
\begin{align}
    [\hat{r}_{j}, \hat{r}_{k}] & = \Omega_{jk}
\end{align}
where $\Omega$ is the symplectic matrix defined by

\begin{align}
    \Omega = \bigoplus_{j=1}^{2N} \begin{pmatrix}
        0 & 1 \\
        -1 & 0
    \end{pmatrix} = \mathbb{I}_{N} \otimes  \begin{pmatrix}
        0 & 1 \\
        -1 & 0
    \end{pmatrix}
\end{align}

where $\mathbb{I}_{N}$ is the $N \times N$ identity matrix.

\subsection{Covariance Matrices}
Given a quantum state $\rho$ of an $N$-mode system, the
corresponding covariance matrix $V$ has elements

\begin{align}
    V_{jk} = \langle \{ \hat{r}_{j}, \hat{r}_{k}\} \rangle -  \langle  \hat{r}_{j} \rangle \langle  \hat{r}_{k} \rangle
\end{align}

where $\{  ., . \}$ denotes the anticommutator and $\langle \hat{A}\rangle $ is the expectation value.
For a valid density matrix $\rho$, the covariance matrix $V$ is real, symmetric, and positive definite ($V > 0$), and it satisfies

\begin{align}
    V + i\Omega \ge 0
\end{align}

Note that the diagonal elements of $V$ represent the variances of the quadrature operators, whereas the non-zero off-diagonal elements correspond to the correlations between different quadratures.

\subsection{Gaussian operations}

Gaussian operations are those that map Gaussian states to Gaussian states, i.e., they preserve Gaussianity. Since a Gaussian state is fully characterised by $\bar{r}$ and $V$, a Gaussian operation is completely specified by its action on $\bar{r}$ and $V$. Unitary Gaussian operations are those generated by Hamiltonians that are at most quadratic in the mode operators $\hat{a}_{j}$ and $\hat{a}^{\dag}_{j}$. These operations correspond precisely to symplectic transformations of both the displacement vector and the covariance matrix.
They map
\begin{align}
    V \to SVS^{T}, \: \text{and} \: \bar{r} \to S\bar{r} + d
\end{align}
where $d$ is a $2N$-dimensional real vector of displacements,
and the matrix $S$ satisfies
\begin{align}
    S\Omega S^{T} = \Omega
\end{align}

Gaussian unitaries are therefore fully described by a symplectic matrix $S$ and a displacement vector $d$. Each Gaussian unitary corresponds to a symplectic transformation, and conversely, for every symplectic transformation, there exists a quadratic Hamiltonian and an associated Gaussian unitary that generates it.

It follows from the Euler decomposition that any Gaussian unitary on $N$ modes can be implemented by a sequence consisting of a passive linear transformation, followed by single-mode squeezing on each mode, and then another passive linear transformation. In the language of quantum optics, this corresponds to a combination of beam splitters, phase shits, and single-mode squeezers.

\subsection{Gaussian Unitaries}
In this section, we present the symplectic transformations corresponding to the Gaussian unitary operations discussed in the main text such as the beamsplitter and phase shifter.

\subsubsection{Phase shifts}

Applying a phase shift of angle $\phi$ to a single mode modifies the mode operators as
\begin{align}
    \hat{a} \to e^{-i \phi} \hat{a}, \: \hat{a}^{\dag} \to e^{-i \phi} \hat{a}^{\dag}
\end{align}
Thus, as a symplectic transformation on a single mode, the phase shift is represented by a rotation
\begin{align}
    S & = \begin{pmatrix}
        \cos{\phi} & \sin{\phi} \\
        -\sin{\phi} & \cos{\phi}
    \end{pmatrix} = R(\phi) \\
    d & = 0
\end{align}
When phase shifts $\phi_{1},\dots, \phi_{N} $ are applied to each mode
of an $N$-mode system, we have

\begin{align}
    S &= \bigoplus_{j=1}^{N} R(\phi_{j}) \\
    d = 0
\end{align}

\subsubsection{Beam splitters}

We consider a beam splitter with transmittivity $\eta$, which acts on the mode operators $\hat{a}{1}$ and $\hat{a}{2}$ of two modes as
\begin{align}
    \begin{pmatrix}
       \hat{a}'_{1}\\
      \hat{a}'_{2}
    \end{pmatrix} & \to  
       \begin{pmatrix}
        \sqrt{\eta} & \sqrt{1- \eta}\\
        \sqrt{1- \eta} & -  \sqrt{\eta}
    \end{pmatrix} \begin{pmatrix}
        \hat{a}_{1}\\
        \hat{a}_{2}
    \end{pmatrix} 
\end{align}

The beam splitter therefore introduces no displacement, but it does mix the mode operators. As a symplectic transformation on two modes, it can be expressed as

\begin{align}
    S  & =  \begin{pmatrix}
        \sqrt{\eta} \mathbb{I}_{2} & \sqrt{1- \eta}  \mathbb{I}_{2}\\
        \sqrt{1- \eta} \mathbb{I}_{2} & -  \sqrt{\eta} \mathbb{I}_{2}
    \end{pmatrix} \\
    d & = 0
\end{align}

When applied to a pair of modes (say $j, k$) within an $N$-mode system, the symplectic matrix equals $S_{\eta}$ on the $jk$-subspace and identity on the remaining modes.
\begin{align}
    S &= \begin{pmatrix}
        S_{\eta} & 0 \\
        0 & \mathbb{I}_{2(N-2)}
    \end{pmatrix} \\
    d & = 0
\end{align}

\subsection{Projection onto the vacuum}
\label{vacdet}

A measurement performed on subsystem $B$ of a joint Gaussian state $\rho_{AB}$, corresponding to a projection onto the vacuum state $\ket{0}_{B}$, leaves subsystem $A$ in a Gaussian state. The covariance matrix of this conditional state of $A$ is related to the covariance matrix of the original joint state.

Consider a joint system $AB$ in a Gaussian state $\rho_{AB}$ with covariance matrix.

\begin{align}
    V_{AB}  = \begin{pmatrix}
        A & C \\
        C^T & B
    \end{pmatrix}  
\end{align}

The conditional state of subsystem A, resulting from projecting subsystem B onto the vacuum state $\ket{0}_{B}$, is characterised by
\begin{align}
     V_{A} = A - C (B + \mathbf{I})^{-1} C^{T}
\end{align}

Note that the projection is taken with respect to a coherent state. Since the covariance matrix projection is independent of the amplitude of the coherent state, it can be equivalently evaluated for the zero-amplitude coherent state (the vacuum). However, this approach does not yield the probability of projecting onto the vacuum state. It only provides the projected covariance matrix.

\section{Number state basis analysis for two-mode case}
\label{appendixA}

We will conduct a step-by-step analysis in the number state basis for the state $\ket{\psi}_{\text{Out}} = \hat{\phi_{1}} \hat{B}(\theta) \hat{\phi_{2}} \ket{\psi_{\text{In}}}$, as outlined in the main text.

In the Heisenberg picture, the annihilation operators are transformed via the linear unitary
Bogoliubov transformation for the two-mode case is,
\begin{align}
    \begin{pmatrix}
       \hat{a}'_{1}\\
      \hat{a}'_{2}
    \end{pmatrix} & \to  
     \begin{pmatrix}
        e^{\phi_{1}} & 0\\
        0 & 1
    \end{pmatrix}  \begin{pmatrix}
        \sqrt{\eta} & \sqrt{1- \eta}\\
        \sqrt{1- \eta} & -  \sqrt{\eta}
    \end{pmatrix} \begin{pmatrix}
        1 & 0\\
        0  & e^{\phi_{2}}
    \end{pmatrix} \begin{pmatrix}
        \hat{a}_{1}\\
        \hat{a}_{2}
    \end{pmatrix} \\
    & = \begin{pmatrix}
        A11 & A12 \\
        A21 & A22
    \end{pmatrix} \begin{pmatrix}
        \hat{a}_{1}\\
        \hat{a}_{2}
    \end{pmatrix} \\
   \hat{a}'_{1} & \to  A11\hat{a}_{1}^{\dag } + A12\hat{a}_{2}^{\dag } \\
  \hat{a}'_{2} & \to  A21\hat{a}_{1}^{\dag } + A22 \hat{a}_{2}^{\dag }\ 
\end{align}

where,
\begin{align}
    A11 &= e^{\phi_{1}} \sqrt{\eta}  \\
A12 & = e^{\phi_{1}}  \sqrt{1- \eta} e^{\phi_{2}}\\
A21 & = \sqrt{1- \eta} \\
A22 &= -\sqrt{\eta} e^{\phi_{2}}
\end{align}
$\eta = \cos^{2}{\theta}$ is the beamsplitter transmissivity and $\phi_{j}$ is the phase angle in mode $j$.

The output state without using UA protocol is,

\begin{align}
    \ket{\psi}_{\text{Out}} = & \hat{U}\ket{\psi_{\text{In}}} \nonumber\\
   = &  \hat{\phi_{1}} \hat{B}(\theta) \hat{\phi_{2}} \ket{\xi_{1}} \ket{\xi_{2}} \nonumber \\
   = &  \hat{\phi_{1}} \hat{B}(\theta) \hat{\phi_{2}} e^{\frac{r_{1}}{2}( e^{-i \phi_{1}} \hat{a}^{2} -  e^{i \phi_{1}} \hat{a}^{\dag 2})} \nonumber \\
 &   e^{\frac{r_{2}}{2}( e^{-i \phi_{2}} \hat{a}^{2} - e^{i \phi_{2}} \hat{a}^{\dag 2})} \ket{0}\ket{0} \nonumber \\
    =  &  (\cosh{r})^{-1} e^{\frac{-\tanh{r}}{2}\big(  \hat{a}^{'\dag }_{1} \big)^{2}}  e^{\frac{-\tanh{r}}{2}\big(  \hat{a}^{'\dag }_{2} \big)^{2}}  \ket{0}\ket{0} \nonumber \\
                            = & (\cosh{r})^{-1} e^{\frac{-\tanh{r}}{2}\big(  A11\hat{a}_{1}^{\dag } + A12\hat{a}_{2}^{\dag } \big)^{2}}  \nonumber \\
                           & e^{\frac{\tanh{r}}{2}\big(A21\hat{a}_{1}^{\dag } + A22 \hat{a}_{2}^{\dag }\big)^{2}} \ket{0}\ket{0} \label{noUA1}
\end{align}

\section{Symplectic map analysis for UA n=2}
\label{appendixB}
The following box outlines the algorithm for implementing the unitary averaging protocol in two-mode Gaussian systems, particularly in the context of Gaussian boson sampling circuits. The procedure involves symplectic map analysis to determine the final covariance matrix, as discussed in \cite{weedbrook2012gaussian, serafini2023quantum}. The box below presents the UA protocol for two-mode Gaussian states, which can be naturally extended to multi-mode systems. We have implemented this procedure numerically for 10-mode systems, with the results presented in the main text. Compared to calculations in the Fock basis, this symplectic map analysis offers a significantly faster computational approach.

We begin with the symplectic matrix corresponding to the vacuum state, represented by the identity matrix. Then, we apply the symplectic transformations associated with single-mode squeezing operations on the first two modes (denoted as $\text{Ssq}(r_1, r_2)$) along with a relative phase shift, indicated by the operator SPhase. Applying these transformations to the vacuum state yields the covariance matrix of the input state: two squeezed modes combined with two vacuum modes. In the Fock basis, this input state is represented as
$ \ket{ \psi_{\text{In}}}  =  \ket{\xi_{1}} \ket{\xi_{2}} \ket{0}\ket{0}$,
as shown in Eq.~\ref{inputFock} of the main text.

Then we express the symplectic form of all the relevant unitaries, including the encoding beamsplitters, noise-inducing beamsplitters, and phase shifters, which are denoted as BSencode, BSerror, and PSerror respectively, in the below box. Using these, we constructed the covariance matrix that results from sequentially applying the encoding operation, the error unitaries, and the decoding step, as shown in Fig.\ref{fig:2MODE2}. The final covariance matrix obtained after the decoding operation is labeled VDecode, as shown in the box below. We then performed a vacuum measurement on VDecode, as described by Eq.\ref{COVGuru} in the main text.
    
\begin{widetext}
\begin{mdframed}
\begin{align*}
    \Theta & = \frac{\pi}{4} \\
    \text{Vacuum state} & = \text{Identity Matrix}[8] \\
    \text{Ssq}(r_{1}, r_{2}) & = \begin{pmatrix}
        r_{1} & 0 & 0 & 0 & 0 & 0 & 0 & 0 \\
        0 &  r_{1} & 0 & 0 & 0 & 0 & 0 & 0 \\
        0 &  0 & r_{2} &  0 & 0 & 0 & 0 & 0  \\
        0 &  0 & 0 &   r_{2} & 0& 0 & 0 & 0  \\
        0 &  0 & 0 &  0 & 1 & 0 & 0 & 0  \\
        0 &  0 & 0 &  0 & 0 & 1 & 0 & 0  \\
        0 &  0 & 0 &  0 & 0 & 0 & 1 & 0  \\
        0 &  0 & 0 &  0 & 0 & 0 & 0 & 1
    \end{pmatrix}\\
    \text{VSMSV} & = \text{Ssq}(r_{1}, r_{2}) * \text{Vacuum state} *(\text{Ssq}(r_{1}, r_{2}))^{T} \\
    \text{SPhase} & = \begin{pmatrix}
        1 & 0 & 0 & 0 & 0 & 0 & 0 & 0 \\
        0 & 1 & 0 & 0 & 0 & 0 & 0 & 0 \\
        0 &  0 & \cos{\pi/2} &  \sin{\pi/2} & 0 & 0 & 0 & 0  \\
        0 &  0 & -\sin{\pi/2} &   \cos{\pi/2} & 0& 0 & 0 & 0  \\
        0 &  0 & 0 &  0 & 1 & 0 & 0 & 0  \\
        0 &  0 & 0 &  0 & 0 & 1 & 0 & 0  \\
        0 &  0 & 0 &  0 & 0 & 0 & 1 & 0  \\
        0 &  0 & 0 &  0 & 0 & 0 & 0 & 1
    \end{pmatrix}\\
   \text{VsqPhase} & =  \text{SPhase}*\text{VSMSV} *( \text{SPhase})^{T}\\
   \text{BSencode} &= \begin{pmatrix}
        \cos{ \Theta} & 0 & 0 & 0 & \sin{\Theta} & 0 & 0 & 0 \\
        0 & \cos{ \Theta}& 0 & 0 & 0 & \sin{\Theta}  & 0 & 0 \\
        0 &  0 & \cos{ \Theta} &  0 & 0 & 0 & \sin{\Theta}& 0  \\
        0 &  0 & 0 &   \cos{ \Theta} & 0& 0 & 0 & \sin{\Theta}  \\
        -\sin{\Theta}  &  0 & 0 &  0 & \cos{ \Theta} & 0 & 0 & 0  \\
        0 &  -\sin{\Theta}  & 0 &  0 & 0 & \cos{ \Theta} & 0 & 0  \\
        0 &  0 & -\sin{\Theta} &  0 & 0 & 0 & \cos{ \Theta} & 0  \\
        0 &  0 & 0 &  -\sin{\Theta} & 0 & 0 & 0 & \cos{ \Theta}
    \end{pmatrix}\\
    \text{Vencode} & = \text{BSencode} * \text{VsqPhase} *(\text{BSencode})^{T}\\
    \text{PSerror} & = \begin{pmatrix}
        \cos{\phi_{1}} &  \sin{\phi_{1}}  & 0 & 0 & 0 & 0 & 0 & 0 \\
        -\sin{\phi_{1}} &  \cos{\phi_{1}} & 0 & 0 & 0 & 0 & 0 & 0 \\
        0 &  0 & \cos{\phi_{2}} &  \sin{\phi_{2}} & 0 & 0 & 0 & 0  \\
        0 &  0 & -\sin{\phi_{2}} &  \cos{\phi_{2}} & 0& 0 & 0 & 0  \\
        0 &  0 & 0 &  0 & \cos{\phi_{3}} &  \sin{\phi_{3}} & 0 & 0  \\
        0 &  0 & 0 &  0 & -\sin{\phi_{3}} &  \cos{\phi_{3}} & 0 & 0  \\
        0 &  0 & 0 &  0 & 0 & 0 & \cos{\phi_{4}} &  \sin{\phi_{4}}   \\
        0 &  0 & 0 &  0 & 0 & 0 & -\sin{\phi_{4}} &  \cos{\phi_{4}}
    \end{pmatrix}\\
    \text{BSerror} & = \begin{pmatrix}
        \cos{\theta_{1}} & 0 & \sin{\theta_{1}} & 0 & 0 & 0 & 0 & 0 \\
        0 &  \cos{\theta_{1}} & 0 & \sin{\theta_{1}} & 0 & 0 & 0 & 0 \\
        -\sin{\theta_{1}} &  0 &\cos{\theta_{1}}&  0 & 0 & 0 & 0 & 0  \\
        0 & -\sin{\theta_{1}} & 0 &   \cos{\theta_{1}}& 0& 0 & 0 & 0  \\
        0 &  0 & 0 &  0 & \cos{\theta_{2}} & 0 & \sin{\theta_{2}} & 0  \\
        0 &  0 & 0 &  0 & 0 & \cos{\theta_{2}} & 0 & \sin{\theta_{2}} \\
        0 &  0 & 0 &  0 & -\sin{\theta_{2}} & 0 & \cos{\theta_{2}} & 0  \\
        0 &  0 & 0 &  0 & 0 & -\sin{\theta_{2}}& 0 & \cos{\theta_{2}}
    \end{pmatrix}\\
    \text{TotError} &= (\text{PSerror}) * (\text{BSerror})\\
    \text{VTotError} & =  \text{TotError} * \text{Vencode} *( \text{TotError})^{T} \\
    \text{VDecode} &= (\text{BSencode})^{T}*\text{VTotError} *\text{BSencode} 
\end{align*}
\text{where `S' represents symplectic matrix and `V' represents covariance matrix.}
\end{mdframed}
\end{widetext}

\end{appendix}

\bibliography{draftGBS}

\end{document}